\documentclass{aa}  

\usepackage{graphicx}
\usepackage{txfonts}
\usepackage[]{hyperref}
\usepackage{breakurl}
\newcommand{\urlwofont}[1]{\urlstyle{same}\url{#1}}
\usepackage{upgreek}

\def\HII{H\,{\sc ii}} 
\def\NaID{Na\,{\sc i}~D} 
\def\HeII{He\,{\sc ii}} 
\def\CI{C\,{\sc i}} 
\def\CII{C\,{\sc ii}} 
\def\CIII{C\,{\sc iii}} 
\def\CaII{Ca\,{\sc ii}} 
\def\OI{O\,{\sc i}} 
\def\FeII{Fe\,{\sc ii}} 
\def\ScII{Sc\,{\sc ii}} 
\def\SiII{Si\,{\sc ii}} 
\def\MgII{Mg\,{\sc ii}}

\begin{document}

   \title{SN 2005at -- A neglected type Ic supernova at 10 Mpc}

   \subtitle{}

   \author{E.~Kankare\inst{1,2}
          \and
          M.~Fraser\inst{3}
          \and
          S.~Ryder\inst{4}
          \and
          C.~Romero-Ca\~nizales\inst{5,6}
          \and
          S.~Mattila\inst{2}
          \and
          R.~Kotak\inst{7}
          \and 
          P.~Laursen\inst{8}
          \and 
          L.~A.~G.~Monard\inst{9}
          \and
          M.~Salvo\inst{10}
          \and
          P.~V\"ais\"anen\inst{11,12}
          }

   \institute{Tuorla Observatory, Department of Physics and Astronomy, University of Turku, V\"ais\"al\"antie 20, 21500 Piikki\"o, Finland\\
              e-mail: erkki.kankare@utu.fi
              \and
              Finnish Centre for Astronomy with ESO (FINCA), University of Turku, V\"ais\"al\"antie 20, 21500 Piikki\"o, Finland
              \and
              Institute of Astronomy, University of Cambridge, Madingley Road, Cambridge CB3 0HA, UK
              \and
              Australian Astronomical Observatory, PO Box 915, North Ryde, NSW 1670, Australia
              \and
              Instituto de Astrof\'{\i}sica, Facultad de F\'{\i}sica, Pontificia Universidad Cat\'olica de Chile, Casilla 306, Santiago 22, Chile
              \and
              Millennium Institute of Astrophysics, Vicu\~na Mackenna 4860, 7820436 Macul, Santiago, Chile
              \and
              Astrophysics Research Centre, School of Mathematics and Physics, Queen's University Belfast, BT7 1NN, UK
              \and
              Dark Cosmology Centre, Niels Bohr Institute, University of Copenhagen, Juliane Maries Vej 30, 2100 Copenhagen, Denmark
              \and
              Bronberg Observatory, Centre for Backyard Astrophysics (Pretoria), 0056 Tiegerpoort, South Africa
              \and
              Research School of Astronomy and Astrophysics, Australian National University, Cotter Road, Weston Creek, ACT 2611, Australia
              \and
              South African Astronomical Observatory, PO Box 9, Observatory 7935, Cape Town, South Africa
              \and
              Southern African Large Telescope, PO Box 9, Observatory 7935, Cape Town, South Africa
             }

   \date{Received 9 July 2014 / Accepted 1 September 2014}

 
  \abstract{We present a photometric and spectroscopic study of a reddened type Ic supernova (SN) 2005at. We report our results based on the available data of SN~2005at, including late-time observations from the \textit{Spitzer} Space Telescope and the \textit{Hubble} Space Telescope. In particular, late-time mid-infrared observations are something rare for type Ib/c SNe. In our study we find SN~2005at to be very similar photometrically and spectroscopically to another nearby type Ic SN~2007gr, underlining the prototypical nature of this well-followed type Ic event. The spectroscopy of both events shows similar narrow spectral line features. The radio observations of SN~2005at are consistent with fast evolution and low luminosity at radio wavelengths. The late-time \textit{Spitzer} data suggest the presence of an unresolved light echo from interstellar dust and dust formation in the ejecta, both of which are unique observations for a type Ic SN. The late-time \textit{Hubble} observations reveal a faint point source coincident with SN~2005at, which is very likely either a declining light echo of the SN or a compact cluster. For completeness we study ground-based pre-explosion archival images of the explosion site of SN~2005at, however this only yielded very shallow upper limits for the SN progenitor star. We derive a host galaxy extinction of $A_{V}\approx1.9$~mag for SN~2005at, which is relatively high for a SN in a normal spiral galaxy not viewed edge-on.}
   

   \maketitle

\section{Introduction}

Core-collapse supernova (CCSN) explosions are highly energetic and terminal events that end the life cycles of some of the most massive stars ($M \gtrsim 8 \mathrm{M}_{\sun}$). In the commonly used SN classification scheme, CCSNe are divided into type II SNe and type Ib/c SNe based on the presence or absence of hydrogen in their spectra, respectively. Type Ib/c SNe are furthermore divided similarly into type Ib and Ic classes based on the presence or absence of helium in their spectra, respectively. For a review of the classification summary, see \citet{filippenko97}. 

Classically, the progenitors of type Ib/c SNe have been thought to be Wolf-Rayet stars that have lost their outer envelopes because of strong stellar winds \citep[see][for a review]{crowther07}. Whereas progenitor detections from pre-explosion high-resolution images have robustly shown type IIP SNe to originate from red supergiants, type Ib/c SN progenitors have evaded confirmed identifications \citep{smartt09a, eldridge13}. Recently, \citet{cao13} have found a progenitor candidate from deep pre-explosion images that is consistent with a single Wolf-Rayet star for the type Ib SN iPTF13bvn. However, based on hydrodynamical modelling of the SN both \citet{fremling14} and \citet{bersten14} have suggested a progenitor with significantly lower mass. Furthermore, \citet{bersten14} performed binary evolution calculations showing that the pre-SN photometry was compatible with a low-mass progenitor. In fact, it is likely that the less massive stars can lose their outer envelopes owing to binary interaction \citep[e.g.][]{podsiadlowski92}, and these systems can form an alternative progenitor channel for type Ib/c SNe. There are also arguments from relative SN rates \citep[e.g.][]{smith11, eldridge13} and SN ejecta masses \citep{lyman14} that a high percentage of stripped envelope SNe come from binary systems. Furthermore, recent statistical studies \citep{anderson12, kangas13} of the association between different CCSN types and the host galaxy emission have suggested similar progenitor masses for type II and type Ib SNe and significantly higher masses for the progenitors of type Ic SNe. 

During the past 20 years only five type Ib/c SNe have been discovered within $\lesssim$10~Mpc: the normal (i.e. non-broad line) but somewhat fast evolving type Ic SN~1994I \citep[e.g.][]{richmond96}, the broad-lined type Ic SN~2002ap \citep[e.g.][]{foley03}, the normal type Ic SN~2005at \citep{martin05}, the normal type Ic SN~2007gr \citep[e.g.][]{hunter09}, and the normal type Ic SN~2012fh \citep{nakano12} classified at a late stage. Discoveries of nearby SNe offer excellent opportunities for detailed multi-wavelength follow-up campaigns and studies of these events. 

Here we study the type Ic SN~2005at and report the available data. In Sect.~\ref{sect:2005at} SN~2005at is introduced, the optical and radio observations are described in Sect.~\ref{sect:obs} and analysed in Sect.~\ref{sect:analysis}. The late-time \textit{Spitzer} Space Telescope and \textit{Hubble} Space Telescope (HST) observations of SN~2005at are investigated in Sect.~\ref{sect:late-time} and the explosion site is studied more closely in Sect.~\ref{sect:site}. Conclusions are given in Sect.~\ref{sect:conclusions}.

\section{SN 2005at in NGC 6744}
\label{sect:2005at}

\citet{martin05} report both the discovery of SN~2005at, located at $\alpha = 19^{\mathrm{h}} 09^{\mathrm{m}} 53\fs62$ and $\delta = -63\degr 49\arcmin 24\farcs1$ (equinox J2000.0), in images obtained on 2005 March 15.83 UT and an independent discovery by Monard on 2005 March 5.139 UT. Unfortunately the comparison images, observed before the SN explosion, do not provide strong constraints on the explosion date of SN~2005at, as they were obtained in 2004. SN~2005at was classified by \citet{schmidt05} with a spectrum obtained on 2005 March 19.77 UT, concluding the SN to be a type Ic event two weeks past maximum light and spectroscopically similar to SN~1994I. The authors also note the red colour of the spectrum.  

\citet{perna08} made use of the \textit{Swift}/XRT X-ray upper limit of SN~2005at ($L_{2-10~\mathrm{keV}} < 3.0 \times 10^{38}$~erg~s$^{-1}$ on 2006 November 31) in their statistical sample of 100 CCSNe to study the intial spin periods of neutron stars. Due to the non-detection in X-ray the nature of the compact remnant (a neutron star or a black hole) is unclear and therefore constraints on the progenitor cannot be placed using these data. 

SN~2005at is located in the nearby southern, grand design spiral galaxy NGC~6744 with an inclination of $50\degr \pm 4\degr$ \citep{ryder99}, and despite its proximity it has not been studied extensively. NGC~6744 is an extended and barred SAB(r)bc type galaxy which hosts a central ring structure \citep{devaucouleurs63}. The galaxy is isolated and the only prominent companion it has is an irregular dwarf galaxy NGC~6744A of IB(s)m type, interacting with NGC~6744 \citep{ryder99}. We also note that no other SN has been reported in this galaxy. 

For the distance of NGC~6744 we adopt $9.5 \pm 0.6$~Mpc \citep[via the Extragalactic Distance Database;\footnote{\urlwofont{http://edd.ifa.hawaii.edu/}}][]{tully09} measured by \citet{jacobs09} with the tip of the red giant branch (TRGB) method \citep{lee93}. The derived distance is based on a HST \textit{F606W}$-$\textit{F814W} vs. \textit{F814W} colour--magnitude diagram of NGC~6744. At this host galaxy distance the projected distance of SN~2005at to the optical nucleus of the NGC~6744, located at $\alpha = 19^{\mathrm{h}} 09^{\mathrm{m}} 46\fs1$ and $\delta = -63\degr 51\arcmin 27\farcs1$ (J2000.0; via NED), corresponds roughly to 6.1~kpc.

\citet{ryder95} studied some of the most luminous \HII\ regions of NGC~6744 and reported super-solar metallicities in the region denoted n6744p1a2, with a projected distance of $\sim$1.5~arcmin ($\sim$4~kpc) from the explosion site of SN~2005at, and at a similar radial distance from the host galaxy centre. Recently, \citet{pilyugin14} studied metallicity abundances of 130 nearby late-type galaxies finding a relatively steep oxygen abundance gradient and a large extrapolated central oxygen abundance $12+\log(\mathrm{O}/\mathrm{H})_{R_{0}}$ (-0.726$\pm$0.060~dex~$R_{25}^{-1}$ and 8.88$\pm$0.03, respectively) in NGC~6744. At the radial distance of SN~2005at this corresponds to an oxygen abundance of roughly 8.7. In their extensive sample of galaxies, compared to NGC~6744, only three galaxies (UGC~02345, NGC~5457, and NGC~4713) had steeper oxygen abundance gradients and only two galaxies (NGC~2841 and NGC~4501) had larger $12+\log(\mathrm{O}/\mathrm{H})_{R_{0}}$ values. Furthermore, the nitrogen abundance gradient and $12+\log(\mathrm{N}/\mathrm{H})_{R_{0}}$ value of NGC~6744 (-1.996$\pm$0.157~dex~$R_{25}^{-1}$ and 8.66$\pm$0.07, respectively) were even more extreme with no other galaxy in the sample of \citet{pilyugin14} having a steeper gradient and only one galaxy (NGC~2841) having a larger central nitrogen abundance. For the radial distance of SN~2005at this suggests a nitrogen abundance of roughly 8.2.

Due to reasons such as high line-of-sight extinction, position on the sky or host galaxy type, some nearby SNe end up being neglected or even completely lost, see e.g. the late-time archival discovery of SN~2008jb \citep{prieto12}. To our knowledge no detailed follow-up campaign of SN~2005at was ever carried out. However, we have made use of the available archival data of the SN to study its nature and report these data here. We expect this to be very useful for statistical studies of SNe, which use the parameters of the nearby SNe for calibration. Due to the small distance of SN~2005at, it has already been included in several studies of CCSN rates \citep{smartt09b, horiuchi11, horiuchi13, botticella12, mattila12} with the aim to make use of a local and as complete as possible, volume-limited sample of CCSNe. In particular, complete studies of nearby CCSNe are critical for deriving reliable extinction corrections for CCSN rates from the local Universe to cosmological distances \citep{dahlen12, mattila12, melinder12}.

\section{Follow-up observations}
\label{sect:obs}

\subsection{Optical imaging}

We carried out an unfiltered imaging follow-up of the SN from the discovery, at the Bronberg observatory with the Meade 30~cm LX 200 telescope and SBIG ST-7XME CCD camera, and reported the preliminary photometry\footnote{\urlwofont{http://www.astrosurf.com/snweb2/2005/05at/05atMeas.htm}} using an \textit{R}-band zeropoint. Unfortunately, most of the original image files, covering photometry over three months from discovery, have been lost and we have only managed to recover three epochs of the data, i.e. March 19.1, April 17.1, and June 18.1. 

We monitored SN~2005at for about three weeks in April 2005, with nightly observations at the Danish 1.54-m telescope in La Silla, Chile with the Danish Faint Object Spectrograph and Camera \citep[DFOSC;][]{andersen95} in \textit{UBVRI}. 

\begin{figure}
\includegraphics[width=\linewidth]{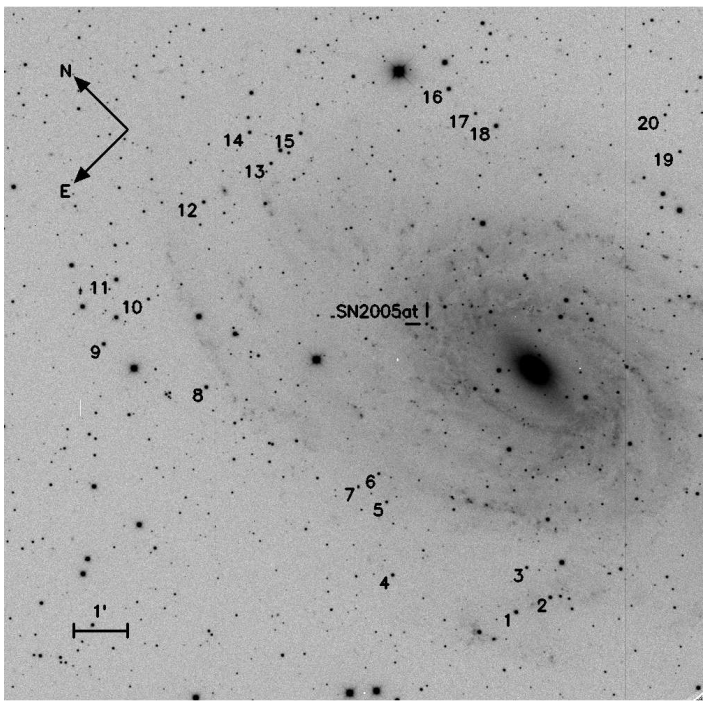}
\caption{13\arcmin$\times$13\arcmin\ DFOSC \textit{V}-band image of the field of SN~2005at observed on 2005 April 10.4 UT. Sequence stars used to calibrate the photometry and the image orientation are indicated.}
\label{fig:field}
\end{figure}

The DFOSC images were reduced using the QUBA\footnote{Python package specifically designed by S. Valenti for SN imaging and spectra reduction. For more details on the pipeline, see \citet{valenti11}.} pipeline. The pipeline makes use of IRAF\footnote{IRAF is distributed by the National Optical Astronomy Observatories, which are operated by the Association of Universities for Research in Astronomy, Inc., under cooperative agreement with the National Science Foundation.}  tasks and the reductions included standard bias subtraction and flat field correction steps. The QUBA pipeline was also used to carry out the photometry of SN~2005at. A selection of photometric nights in La Silla were identified based on multiple epochs of photometric standard star field \citep{landolt92} observations. The standard star observations in photometric conditions were used to derive instrumental zeropoints for DFOSC and from those, magnitudes of the sequence stars indicated in Fig.~\ref{fig:field} of the field of SN~2005at, and listed in Table~\ref{table:seq}, were calculated. The standard star observations were also used to derive the average colour terms for the DFOSC filters used. The photometry of the SN was carried out in the pipeline as point spread function (PSF) fitting photometry, making use of the {\sc daophot} tasks in IRAF. The PSF shape in the images is derived from the local sequence stars. The photometry includes simultaneous fitting of the SN and possible nearby stars contaminating the photometry, fitting the background (and foreground) at the SN line-of-sight, and aperture correction. We note that for the \textit{U} band in particular, careful background fitting and nearby source fitting was crucial, as the SN exploded close to a blue extended source, likely a nearby \HII\ region. The error on the SN photometry is estimated to be the standard deviation of photometry of nine artificial stars placed in the image in different locations near the SN position. The reported total error is the quadrature sum of this standard deviation and the standard error of the mean of the sequence star zeropoint.

\begin{table*}
\caption{Sequence star magnitudes derived for the NGC~6744 field.}
\centering
\begin{tabular}{ccccccccc}
\hline
\hline
Star & $\alpha$ & $\delta$ & $m_{U}$ & $m_{B}$ & $m_{V}$ & $m_{R}$ & $m_{I}$\\
\# & (J2000.0) & (J2000.0) & (mag) & (mag) & (mag) & (mag) & (mag)\\ 
\hline
 1 & $19^{\mathrm{h}}10^{\mathrm{m}}17\fs943$ & $-63\degr 54\arcmin 26\farcs41$ & 16.63 (0.05) & 15.89 (0.01) & 14.97 (0.02) & 14.44 (0.01) & 13.99 (0.01) \\
 2 & $19^{\mathrm{h}}10^{\mathrm{m}}12\fs027$ & $-63\degr 54\arcmin 42\farcs29$ & 16.99 (0.03) & 16.29 (0.01) & 15.39 (0.02) & 14.87 (0.02) & 14.43 (0.02) \\
 3 & $19^{\mathrm{h}}10^{\mathrm{m}}11\fs280$ & $-63\degr 53\arcmin 59\farcs76$ & 16.98 (0.08) & 16.95 (0.01) & 16.23 (0.02) & 15.78 (0.01) & 15.33 (0.01) \\
 4 & $19^{\mathrm{h}}10^{\mathrm{m}}28\fs326$ & $-63\degr 52\arcmin 18\farcs26$ & 17.08 (0.04) & 16.66 (0.02) & 15.87 (0.03) & 15.45 (0.02) & 14.99 (0.05) \\
 5 & $19^{\mathrm{h}}10^{\mathrm{m}}20\fs079$ & $-63\degr 51\arcmin 14\farcs35$ & 18.83 (0.03) & 17.58 (0.04) & 16.49 (0.03) & 15.91 (0.04) & 15.30 (0.12) \\
 6 & $19^{\mathrm{h}}10^{\mathrm{m}}17\fs546$ & $-63\degr 50\arcmin 45\farcs71$ & 17.18 (0.05) & 16.99 (0.09) & 16.37 (0.03) & 16.04 (0.04) & 15.53 (0.15) \\
 7 & $19^{\mathrm{h}}10^{\mathrm{m}}21\fs727$ & $-63\degr 50\arcmin 39\farcs33$ & 17.24 (0.09) & 16.75 (0.05) & 15.89 (0.03) & 15.45 (0.04) & 15.00 (0.05) \\
 8 & $19^{\mathrm{h}}10^{\mathrm{m}}27\fs958$ & $-63\degr 47\arcmin 16\farcs47$ & 16.26 (0.07) & 15.87 (0.05) & 15.10 (0.03) & 14.72 (0.02) & 14.34 (0.03) \\
 9 & $19^{\mathrm{h}}10^{\mathrm{m}}35\fs005$ & $-63\degr 45\arcmin 19\farcs39$ & 16.34 (0.05) & 15.89 (0.02) & 15.06 (0.03) & 14.66 (0.02) & 14.29 (0.02) \\
 10 & $19^{\mathrm{h}}10^{\mathrm{m}}24\fs189$ & $-63\degr 45\arcmin 19\farcs34$ & 16.82 (0.04) & 16.84 (0.02) & 16.24 (0.03) & 15.90 (0.02) & 15.55 (0.03) \\
 11 & $19^{\mathrm{h}}10^{\mathrm{m}}25\fs676$ & $-63\degr 44\arcmin 37\farcs74$ & 15.82 (0.04) & 15.53 (0.02) & 14.80 (0.03) & 14.42 (0.02) & 14.07 (0.01) \\
 12 & $19^{\mathrm{h}}10^{\mathrm{m}}05\fs739$ & $-63\degr 44\arcmin 45\farcs85$ & 16.62 (0.06) & 16.40 (0.03) & 15.66 (0.02) & 15.28 (0.02) & 14.92 (0.03) \\
 13 & $19^{\mathrm{h}}09^{\mathrm{m}}52\fs937$ & $-63\degr 45\arcmin 08\farcs49$ & 17.09 (0.05) & 16.76 (0.02) & 15.99 (0.01) & 15.56 (0.01) & 15.12 (0.02) \\
 14 & $19^{\mathrm{h}}09^{\mathrm{m}}51\fs703$ & $-63\degr 44\arcmin 26\farcs23$ & 16.79 (0.06) & 16.41 (0.02) & 15.66 (0.01) & 15.26 (0.02) & 14.87 (0.02) \\
 15 & $19^{\mathrm{h}}09^{\mathrm{m}}45\fs663$ & $-63\degr 45\arcmin 08\farcs10$ & 16.31 (0.05) & 16.23 (0.02) & 15.63 (0.01) & 15.29 (0.02) & 14.92 (0.01) \\
 16 & $19^{\mathrm{h}}09^{\mathrm{m}}22\fs278$ & $-63\degr 46\arcmin 31\farcs76$ & 15.11 (0.06) & 15.14 (0.01) & 14.68 (0.02) & 14.39 (0.03) & 13.98 (0.04) \\
 17 & $19^{\mathrm{h}}09^{\mathrm{m}}22\fs044$ & $-63\degr 47\arcmin 13\farcs12$ & 16.96 (0.07) & 16.98 (0.01) & 16.46 (0.02) & 16.14 (0.03) & 15.74 (0.03) \\
 18 & $19^{\mathrm{h}}09^{\mathrm{m}}21\fs069$ & $-63\degr 47\arcmin 39\farcs66$ & 15.34 (0.06) & 15.17 (0.02) & 14.51 (0.03) & 14.14 (0.04) & 13.69 (0.05) \\
 19 & $19^{\mathrm{h}}09^{\mathrm{m}}01\fs975$ & $-63\degr 50\arcmin 26\farcs97$ & 17.70 (0.04) & 17.10 (0.02) & 16.27 (0.01) & 15.83 (0.03) & 15.45 (0.08) \\
 20 & $19^{\mathrm{h}}08^{\mathrm{m}}59\fs299$ & $-63\degr 49\arcmin 45\farcs41$ & 17.29 (0.07) & 17.25 (0.01) & 16.66 (0.01) & 16.35 (0.05) & 15.93 (0.03) \\
\hline
\end{tabular}
\tablefoot{The errors are given in brackets. Coordinates are taken from the USNO-B1.0 Catalog \citep{monet03}.}
\label{table:seq}
\end{table*}

In a similar way to the DFOSC images, we carried out with the QUBA pipeline PSF-fitting photometry of the available unfiltered Bronberg observatory images, calibrating them to the \textit{R}-band magnitudes. With the April 17.1 image we find good agreement with the DFOSC \textit{R}-band photometry and when comparing the photometry of the three measured epochs of Bronberg data, we find a systematic offset of roughly $+0.1$~mag compared to the reported measurements. Such a systematic offset is likely to be explained by host galaxy background, arising from the spiral arm structure, contaminating the photometry of the SN. For epochs where the photometry was not re-visited, we apply this systematic shift to the initial measurements and estimate the error of photometry to be roughly $\pm 0.1$~mag. After the correction of the reported photometry, we find the Bronberg data to be quite consistent with our DFOSC photometry over the whole period of overlap. 

We also recovered SN~2005at from late-time \textit{BVI} images of NGC~6744 observed with the 10-m Southern African Large Telescope (SALT) using the SALT imaging and acquisition camera \citep[SALTICAM;][]{odonoghue06}, obtained as part of the first-light observations of the telescope. Photometry was again carried out with the QUBA pipeline. 

\begin{figure}
\includegraphics[width=\linewidth]{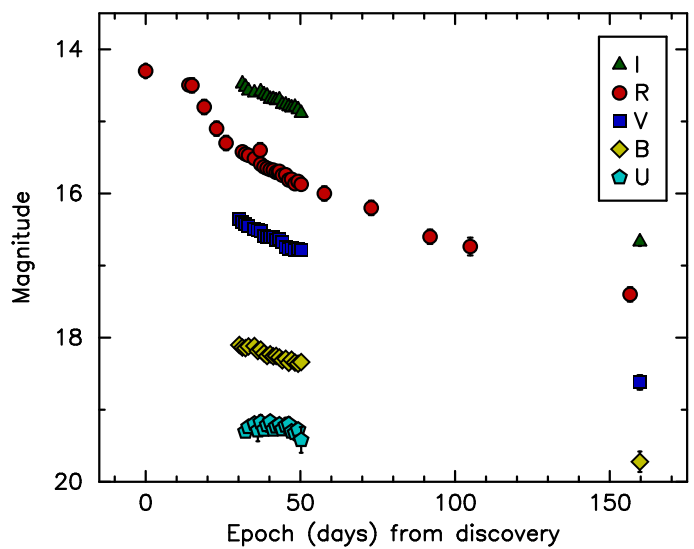}
\caption{Observed \textit{UBVRI} light curves of SN~2005at, suggesting a relatively high host galaxy extinction based on the red colour of the SN. The epoch is set based on the earliest detection of the SN.}
\label{fig:obs_lc}
\end{figure}

The final photometry is reported in Table~\ref{table:phot_05at}, and presented in Fig.~\ref{fig:obs_lc}. The red colour of the SN suggests relatively high host galaxy extinction and this is studied further in Sect.~\ref{sec:analysis_phot}.

\begin{table*}
\caption{Optical photometry for SN~2005at. The errors are given in brackets.}
\centering
\begin{tabular}{cccccccc}
\hline
\hline
JD & $m_{U}$ & $m_{B}$ & $m_{V}$ & $m_{R}$ & $m_{I}$ & Telescope\\
(2400000+) & (mag) & (mag) & (mag) & (mag) & (mag) & \\ 
\hline
53434.64  & $\ldots$ & $\ldots$ & $\ldots$ & 14.3 (0.1) & $\ldots$ & Bronberg\\
53448.63 & $\ldots$ & $\ldots$ & $\ldots$ & 14.50 (0.03) & $\ldots$ & Bronberg\\
53449.61  & $\ldots$ & $\ldots$ & $\ldots$ & 14.5 (0.1) & $\ldots$ & Bronberg\\
53453.57  & $\ldots$ & $\ldots$ & $\ldots$ & 14.8 (0.1) & $\ldots$ & Bronberg\\
53457.56  & $\ldots$ & $\ldots$ & $\ldots$ & 15.1 (0.1) & $\ldots$ & Bronberg\\
53460.64  & $\ldots$ & $\ldots$ & $\ldots$ & 15.3 (0.1) & $\ldots$ & Bronberg\\
53464.91 & $\ldots$ & 18.10 (0.02) & 16.35 (0.02) & $\ldots$ & $\ldots$  & DFOSC\\
53465.91 & $\ldots$ & 18.14 (0.02) & 16.39 (0.02) & 15.42 (0.02) & 14.48 (0.03) & DFOSC\\
53466.90 & 19.30 (0.06) & 18.15 (0.02) & 16.43 (0.01) & 15.45 (0.01) & 14.52 (0.02) & DFOSC\\
53467.89 & 19.24 (0.07) & 18.12 (0.04) & 16.45 (0.02) & 15.47 (0.01) & 14.57 (0.03) & DFOSC\\
53469.81 & 19.19 (0.04) & 18.12 (0.02) & 16.49 (0.03) & 15.51 (0.02) & 14.60 (0.03) & DFOSC\\
53470.91 & 19.29 (0.15) & 18.19 (0.02) & 16.51 (0.01) & $\ldots$ & $\ldots$ & DFOSC\\
53471.65  & $\ldots$ & $\ldots$ & $\ldots$ & 15.4 (0.1) & $\ldots$ & Bronberg\\
53471.82 & 19.17 (0.10) & 18.17 (0.05) & 16.52 (0.02) & 15.59 (0.01) & 14.59 (0.01) & DFOSC\\
53472.89 & 19.28 (0.05) & 18.22 (0.02) & 16.59 (0.01) & 15.63 (0.01) & 14.63 (0.02) & DFOSC\\
53473.90 & 19.21 (0.09) & 18.26 (0.02) & 16.61 (0.01) & 15.65 (0.01) & 14.65 (0.01) & DFOSC\\
53474.90 & 19.17 (0.08) & 18.23 (0.03) & 16.61 (0.01) & 15.67 (0.02) & 14.69 (0.01) & DFOSC\\
53475.90 & 19.28 (0.04) & 18.26 (0.02) & 16.61 (0.01) & 15.68 (0.01) & 14.69 (0.02) & DFOSC\\
53476.82 & 19.23 (0.07) & 18.25 (0.02) & 16.65 (0.01) & 15.71 (0.02) & 14.71 (0.03) & DFOSC\\
53477.55 & $\ldots$ & $\ldots$ & $\ldots$ & 15.71 (0.05) & $\ldots$ & Bronberg\\
53477.85 & 19.21 (0.05) & 18.28 (0.02) & 16.64 (0.02) & 15.69 (0.03) & 14.70 (0.03) & DFOSC\\
53478.81 & 19.27 (0.06) & 18.32 (0.02) & 16.68 (0.02) & 15.75 (0.02) & 14.76 (0.03) & DFOSC\\
53479.91 & 19.22 (0.11) & 18.29 (0.02) & 16.74 (0.03) & 15.74 (0.02) & 14.77 (0.02) & DFOSC\\
53480.90 & 19.20 (0.09) & 18.35 (0.05) & 16.77 (0.02) & 15.81 (0.01) & 14.79 (0.02) & DFOSC\\
53481.80 & 19.30 (0.08) & 18.31 (0.03) & 16.75 (0.02) & 15.80 (0.01) & 14.81 (0.02) & DFOSC\\
53482.91 & 19.32 (0.10) & 18.35 (0.03) & 16.78 (0.02) & 15.87 (0.01) & 14.80 (0.02) & DFOSC\\
53483.91 & 19.28 (0.08) & 18.36 (0.02) & 16.77 (0.02) & 15.83 (0.01) & 14.83 (0.03) & DFOSC\\
53484.91 & 19.42 (0.18) & 18.34 (0.05) & 16.79 (0.02) & 15.87 (0.02) & 14.89 (0.02) & DFOSC\\
53492.41  & $\ldots$ & $\ldots$ & $\ldots$ & 16.0 (0.1) & $\ldots$ & Bronberg\\
53507.55  & $\ldots$ & $\ldots$ & $\ldots$ & 16.2 (0.1) & $\ldots$ & Bronberg\\
53526.56  & $\ldots$ & $\ldots$ & $\ldots$ & 16.6 (0.1) & $\ldots$ & Bronberg\\
53539.56 & $\ldots$ & $\ldots$ & $\ldots$ & 16.74 (0.12) & $\ldots$ & Bronberg\\
53591.22  & $\ldots$ & $\ldots$ & $\ldots$ & 17.4 (0.1) & $\ldots$ & Bronberg\\
53594.39 & $\ldots$ & 19.72 (0.14) & 18.62 (0.10) & $\ldots$ & 16.67 (0.06) & SALT\\
\hline
\end{tabular}
\label{table:phot_05at}
\end{table*}

\subsection{Spectroscopy}

The classification spectrum of SN~2005at \citep{schmidt05} was obtained on $\mathrm{JD} = 2453449.27$ with the Dual-Beam Spectrograph \citep[DBS;][]{rodgers88} at the Australian National University (ANU) 2.3 metre telescope, located at the Siding Spring Observatory. The observations consisted of 300~s exposures obtained with both the blue and the red arm of the instrument using gratings B300 and R316, respectively. The spectra were reduced using standard methods, wavelength calibrated with internal NeAr lamp spectra, and flux calibrated with spectra of the spectroscopic standard EG~131 \citep{bessell99}. The combined spectrum covers the whole optical region with a wavelength range of 3300$-$10200~\AA, and is discussed further in Sect.~\ref{sec:analysis_spect}. To our knowledge this is the only spectrum obtained of SN~2005at.

\subsection{Radio observations}

Two epochs of radio observations for SN~2005at were obtained at the Australia Telescope Compact Array (ATCA). Over the course of 9 h on 2005 March 30 UT (JD~$=2453460$) four of the six antennas were used in the 6A array configuration to observe SN~2005at, with frequent switching between simultaneous measurements at either 1.38 and 2.37~GHz, or at 4.79 and 8.64~GHz, using a bandwidth of 128~MHz split into 32 channels. A further 7 h of observation using all six antennas on 2005 Apr. 13 UT  (JD~$=2453474$) was obtained, but only at 4.79 and 8.64~GHz. Being just 3 degrees away from SN~2005at, the ATCA primary flux calibrator B1934-638 was able to be used to determine both gain and phase variations with time, as well as provide a bandpass calibration.

The uv-plane data for each frequency band and epoch were flagged for outliers automatically then calibrated using tasks in the Miriad package \citep{sault95}. For the imaging robust weighting was employed to deliver the best compromise between the minimal sidelobes produced by uniform weighting, and the minimal noise achieved with natural weighting. The images were cleaned to an rms noise level $\sim$0.05~mJy at the higher frequencies, and $\sim$0.15~mJy at the lower frequencies. The {\sc imfit} task was used to fit a point source at the expected location of SN~2005at, yielding the fluxes (or upper limits) in Table~\ref{table:radio}. The uncertainties in Table~\ref{table:radio} come from the quadrature sum of the image rms and a fractional error on the absolute flux scale in each band \citep{weiler11}.

\begin{table}
\setlength{\tabcolsep}{4pt}
\caption{ATCA radio observations and 3$\sigma$ upper limits for SN~2005at.}
\centering
\begin{tabular}{cccccccc}
\hline
\hline
JD & 1.38~GHz & 2.37~GHz & 4.79~GHz & 8.64~GHz \\
(2400000+) & (mJy) & (mJy) & (mJy) & (mJy) \\ 
\hline
53460 & 1.10(0.14) & 0.48(0.14) & <0.24 & <0.33 \\
53474 & $\ldots$ & $\ldots$ & <0.33 & <0.33 \\
\hline
\end{tabular}
\tablefoot{The errors are given in brackets.}
\label{table:radio}
\end{table}

\section{Analysis}
\label{sect:analysis}

\subsection{Optical photometry}
\label{sec:analysis_phot}

\begin{figure}
\includegraphics[width=\linewidth]{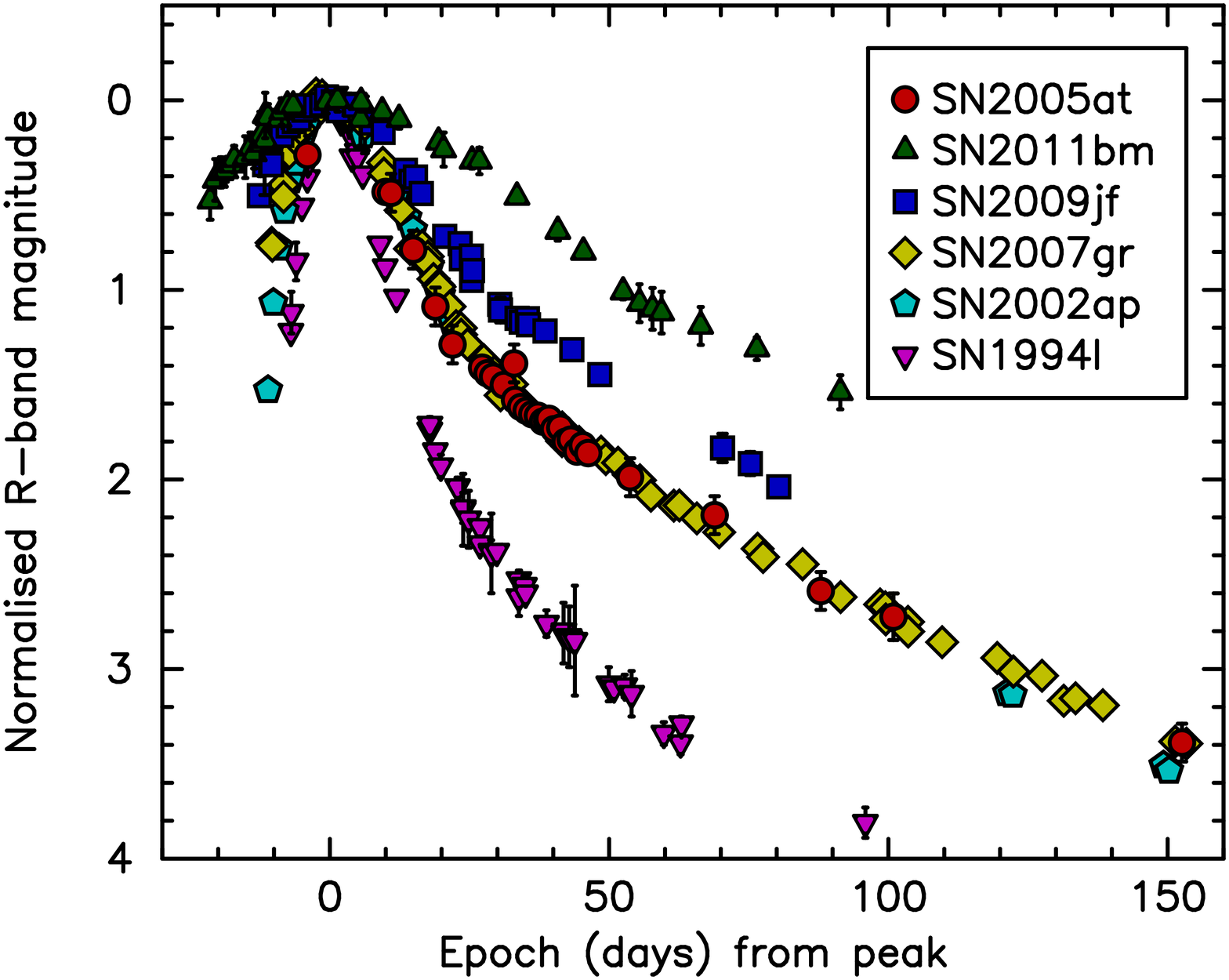}
\caption{\textit{R}-band light curve of SN~2005at compared to those of SNe~1994I \citep[Ic;][]{richmond96}, 2002ap \citep[Ic;][]{foley03}, 2007gr \citep[Ic;][]{hunter09}, 2009jf \citep[Ib;][]{valenti11} and 2011bm \citep[Ic;][]{valenti12}. The data has been shifted such that the SNe have the same peak magnitude. The light curve of SN~2005at shows a very normal photometric evolution and demonstrates significant similarity with SN~2007gr. The epoch is set based on the estimated \textit{R}-band peak of the SNe.}
\label{fig:R_lc}
\end{figure}

In Fig.~\ref{fig:R_lc} the peak normalised \textit{R}-band light curve of SN~2005at is shown in comparison with a selection of other type Ib/c SNe with good photometric coverage. The figure immediately reveals a very similar light curve shape between SNe~2005at and 2007gr, which is a very well followed up nearby type Ic SN, see \citet{hunter09} for an extensive follow-up data set. Recently \citet{chen14} also reported their follow-up observations of SN~2007gr. From the comparison it is also evident that the light curve of the broad-lined type Ic SN~2002ap is similar to that of SN~2005at. In fact, the photometric evolution of SN~2005at is quite typical of a type Ic SN, and it clearly does not belong to the class of very slowly evolving events, such as SN~2011bm. Although SN~2005at was spectroscopically classified to be SN~1994I-like, the light curve of the event shows it not to evolve as rapidly as SN~1994I.

To estimate the line-of-sight host galaxy extinction of SN~2005at from the observed light curves, we apply a similar analysis to that recently presented in \citet{kankare12,kankare14}. In the applied method, the observed light curves of SN~2005at are simultaneously compared to those of well-sampled reference SNe that are assumed to be intrinsically similar to SN~2005at. By minimising the $\chi^{2}$ value three free parameters are derived for SN~2005at via the comparison, namely host galaxy extinction $A_{V}$, the discovery epoch $t_{0}$, and a constant $C$ describing the difference in absolute magnitude between the two SNe in all the bands. If the comparison SN has a very low and/or well-defined line-of-sight extinction this is a very useful approach when studying SNe with relatively high line-of-sight extinctions. In particular, the method is superior compared to approaches making use of empirical relations between the equivalent width of the \NaID\ absorption features and line-of-sight extinction, especially when only low-resolution spectra are available, which is known to be problematic for high extinctions \citep[e.g.][]{poznanski11, phillips13}.

The random photometric errors of the light curves used contribute to the $\chi^{2}$ value of the fit, however the applied method is not very sensitive to these errors when the comparison light curve is well sampled. Furthermore, the SN~2005at data points are weighted based on their error from the $\chi^{2}$ fit. The systematic errors in the light curve arise from a combination of uncertainties in the explosion date, line-of-sight extinction, and distance of the comparison SN, which directly affect the derived parameters $t_{0}$, $A_{V}$, and $C$, respectively, and are taken into account. We specifically note that the errors in the distances both to SN~2005at and to the comparison SN affect only the constant $C$ and do not affect the extinction estimate.

Due to the very similar \textit{R}-band evolution, the well-sampled light curves of the nearby type Ic SN~2007gr were used as a template in the multi-band comparison. Both SNe~2005at and 2007gr exploded in seemingly normal spiral galaxies and we consider in our analysis only the \citet{cardelli89} extinction law with $R_{V}=3.1$, which is widely accepted to describe well the dust reddening conditions in normal galaxies. We adopt a Galactic extinction of $A_{V}=0.118$~mag from \citet{schlafly11} for SN~2005at. For SN~2007gr we adopt, for consistency, from \citet{hunter09} a total line-of-sight extinction of $A_{V}=0.29 \pm 0.06$~mag (based on \NaID\ absorption lines) and a distance of $9.29 \pm 0.69$~Mpc (based on Cepheid distance of another galaxy belonging to the same group). The comparison yields a host galaxy extinction of $A_{V} = 1.9 \pm 0.1$~mag for SN~2005at; that the earliest detection of SN~2005at on 2005 March 5.139 UT took place $4^{+2}_{-4}$ days before the estimated \textit{R}-band maximum; and indicates that SN~2005at was very similar in intrinsic brightness to SN~2007gr, see Table~\ref{table:chi2}. The reported errors are 1$\sigma$ errors inferred from the reduced $\chi^{2}$ probability distributions for the derived parameters, quadrature summed with the errors of SN~2007gr. In particular, within the errors in the distances of the host galaxies of SNe~2005at and 2007gr, the SNe could have been identical in their absolute brightness. The best fit suggests that SN~2005at was slightly brighter than SN~2007gr, however we also note that \citet{schmidt94} derived a slightly higher distance of 10.6$^{+1.9}_{-1.1}$~Mpc for SN~1969L which also exploded in NGC~1058, the host of SN~2007gr, using the expanding photosphere method (EPM) of measuring distances for type II SNe. The derived host galaxy extinction of SN~2005at is consistent within the errors with the preliminary estimate presented in \citet{mattila12}.

\begin{figure}
\includegraphics[width=\linewidth]{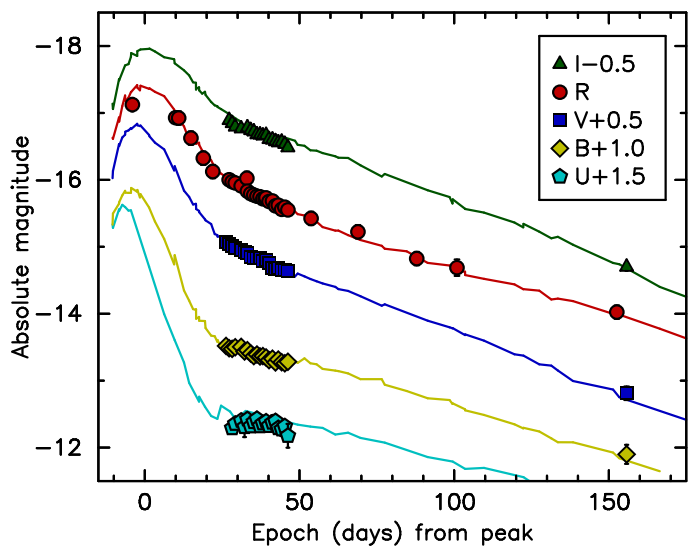}
\caption{Absolute \textit{UBVRI} light curves of SN~2005at, corrected for the derived host galaxy extinction of $A_V=1.9$~mag, compared to those of SN~2007gr (lines). The light curves of SN~2007gr have been vertically shifted by 0.09~mag for a match. In fact, the comparison yields a good match between the two SNe over all the available wavebands and over the whole time range with obtained data. The $\tilde{\chi}^{2} = 5.3$ of the fit is small, though not unity. It is likely that the systematic differences in the \textit{I} and \textit{U} bands dominate the $\tilde{\chi}^{2}$ value, reflecting the small intrinsic differences between the SNe. The epoch is set based on the estimated peak of the \textit{R}-band light curve, similar to Fig.~\ref{fig:R_lc}.}
\label{fig:abs_lc}
\end{figure}

\begin{figure}
\includegraphics[width=\linewidth]{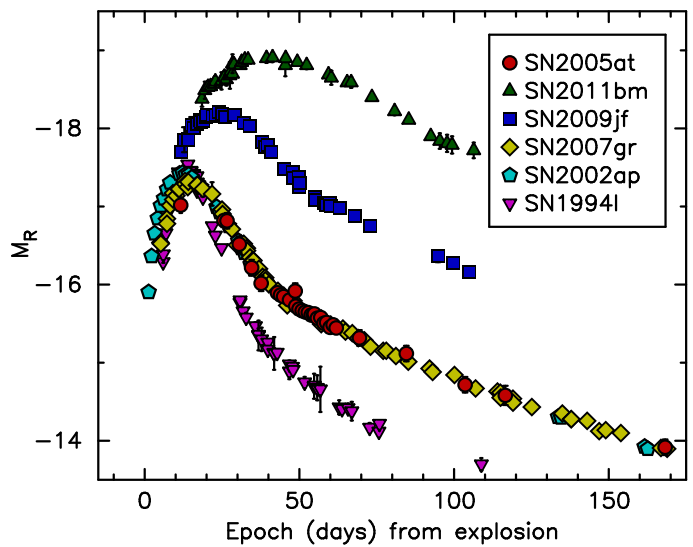}
\caption{Absolute \textit{R}-band light curves of the SNe from Fig.~\ref{fig:R_lc}. The epoch is set from the estimated explosion date.}
\label{fig:comp_lc}
\end{figure}

Assuming for SN~2005at a similar rise time to the \textit{R}-band peak as concluded for SN~2007gr ($15.5 \pm 3.2$ days), the comparison suggests an explosion date of JD~$2453423 \pm 4$, i.e., roughly 2005 February 21. In Fig.~\ref{fig:abs_lc} the extinction corrected absolute light curves of SN~2005at are shown, overlaid on the light curves of SN~2007gr which are shifted with the constant $C$. The similarity between SNe~2005at and 2007gr is obvious and is further emphasized in the comparison to the absolute light curves of other spectroscopically normal type Ib/c SNe in Fig.~\ref{fig:comp_lc}.

 \begin{table}
\caption{Parameters derived for SN~2005at based on the $\chi^{2}$ fit.}
\centering
\begin{tabular}{ccccc}
\hline
\hline
Reference & $A_{V}$ & $t_{0}$ & $C$ & $\tilde{\chi}^{2}$\\
 & (mag) & (days) & (mag) & \\ 
\hline
\vspace{-0.13in}\\
SN 2007gr & $1.9 \pm 0.1$ & $-4^{+2}_{-4}$ & $-0.1^{+0.3}_{-0.2}$ & 5.3 \\
\vspace{-0.13in}\\
\hline
\end{tabular}
\label{table:chi2}
\end{table}

A least-squares fit for the well-sampled DFOSC data, 42 to 62 days from the explosion, yields decline rates of $\gamma_{U} = $ 0.003 $\pm$ 0.003~mag~d$^{-1}$, $\gamma_{B} = $ 0.0134 $\pm$ 0.0009~mag~d$^{-1}$, $\gamma_{V} = $ 0.0214 $\pm$ 0.0006~mag~d$^{-1}$, $\gamma_{R} = $ 0.0235 $\pm$ 0.0006~mag~d$^{-1}$, and $\gamma_{I} = $ 0.0195 $\pm$ 0.0008~mag~d$^{-1}$. The \textit{U}-band decline rate is consistent with there being no decline within the errors. The \textit{U}-band light curve of SN~2007gr appears to be also quite flat in the same phase. The decline rates of the other optical bands are steeper than the expected decline rate of 0.0098~mag~d$^{-1}$ assuming complete $\gamma$-ray and positron trapping in the ejecta from the radioactive decay of $^{56}$Co to $^{56}$Fe. However, this is typical of type Ib/c SNe and furthermore, the epoch range of the DFOSC data covers likely at least partly a transitional phase when the light curve is not yet dominated by the decay of $^{56}$Co, but has still significant contribution from the decay process of $^{56}$Ni to $^{56}$Co. 

With a simple model applied to the bolometric light curve of SN~2007gr \citet{hunter09} inferred explosion parameters, $^{56}$Ni mass, ejecta mass and kinetic energy, for the SN, i.e. $M_{\mathrm{Ni}} = 0.076 \pm 0.020$~$M_{\sun}$, $M_{\mathrm{ej}} = $~2.0$-$3.5~$M_{\sun}$, $E_{\mathrm{k}} = $~1$-$4~$\times 10^{51}$~erg. \citet{mazzali10} carried-out more detailed modelling using the late-time nebular spectra of SN~2007gr finding results consistent with those of \citet{hunter09} and proposing a $M_{\mathrm{ZAMS}} \approx 15$~$M_{\sun}$ progenitor star for the SN. The available follow-up data of SN~2005at is mainly optical and both the light curve evolution and the spectral features (Sect.~\ref{sec:analysis_spect}) are almost identical to those of SN~2007gr. Such a strong similarity, even though the rise of the light curve of SN~2005at is very poorly defined, suggests that these SNe had very similar properties and that SN~2005at likely falls into the above mentioned parameter range. 

\subsection{Classification spectrum}
\label{sec:analysis_spect}

\begin{figure}
\includegraphics[width=\linewidth]{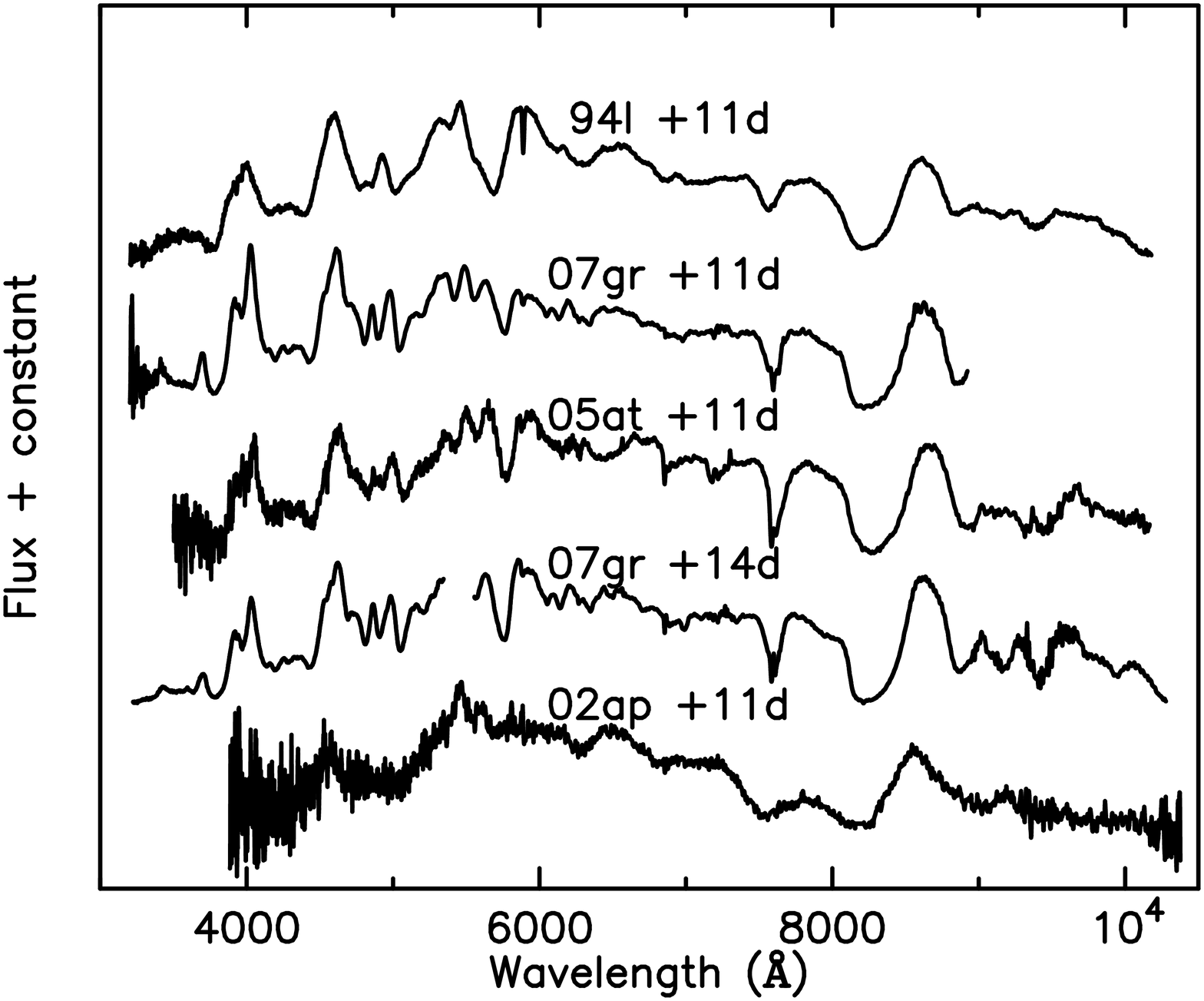}
\caption{Optical spectrum of SN~2005at shown in comparison with the spectra of the type Ic SNe~1994I \citep{filippenko95}, 2002ap \citep{foley03} and 2007gr \citep{hunter09, crockett08}. The spectra are dereddened and the wavelengths corrected to the host galaxy rest frame. Furthermore they are arbitrarily scaled and shifted for clarity. For SN~2005at $A_{V}=1.9$~mag is assumed based on the light curve analysis. Epochs are given relative to the estimated \textit{R}-band maxima of the SNe. The comparison spectra have been obtained from the Weizmann interactive supernova data repository \citep{yaron12}.}
\label{fig:spect}
\end{figure}

Based on the light curve analysis, the classification spectrum was obtained roughly 11~days after the \textit{R}-band peak. Assuming similar light curve evolution for SNe~2005at and 2007gr this corresponds to roughly 15 days after the \textit{B}-band peak and 26 days after the explosion. The spectrum shown in Fig.~\ref{fig:spect} is dereddened based on the light curve analysis and wavelengths rest-frame corrected with a redshift $z=0.002805$ \citep{koribalski04}, and is compared to a selection of stripped-envelope SNe at similar epochs. The spectrum of SN~2005at shares significant similarity in particular with that of SN~2007gr over the whole optical wavelength range. Although the light curve evolution of SN~2002ap shares similarity to that of SN~2005at, the spectroscopic comparison shows a clear difference between the two SNe with broader spectral features arising from SN~2002ap. 

\begin{figure}
\includegraphics[width=\linewidth]{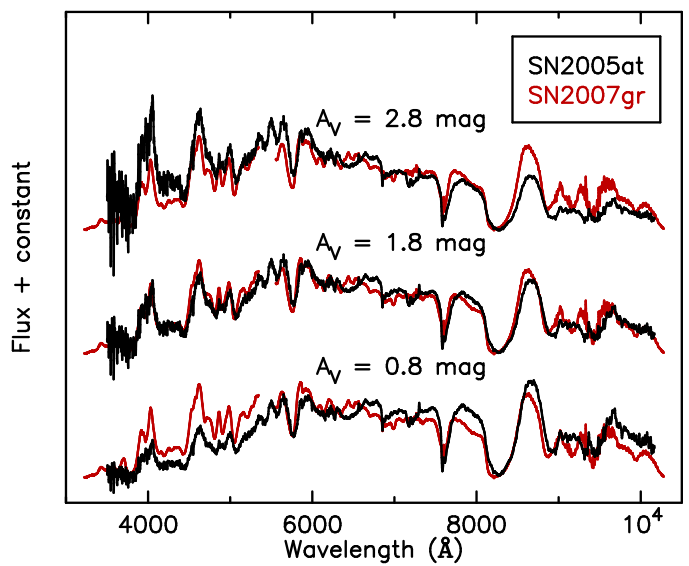}
\caption{Optical spectrum of SN~2005at (black) dereddened with different values of host galaxy extinction, overlaid with the spectra of SN~2007gr (red) obtained 14 days after the \textit{R}-band peak \citep{crockett08}. The best match in the spectroscopic comparison is achieved with SN~2005at host galaxy extinction of $A_{V}=1.8$~mag, which is consistent with the adopted $A_{V}=1.9 \pm 0.1$~mag derived based on the light curves. The spectra are scaled and shifted for clarity.}
\label{fig:comp_spect}
\end{figure}

Based on similar light curve evolution of SNe~2005at and 2007gr, we carry out an additional extinction estimation comparison between the classification spectrum of SN~2005at and an optical spectrum of SN~2007gr \citep{crockett08} obtained at a similar epoch, $\sim$14 days after the \textit{R}-band peak. The comparison follows the approach used in \citet{kankare14} and infers host galaxy extinction for SN~2005at by minimising the standard deviation of the subtraction between the scaled and reddened spectrum of SN~2005at and that of SN~2007gr. The spectroscopic comparison over the full optical range yields the best match with $A_{V}=1.8$~mag, in good agreement with the adopted value $A_{V}=1.9 \pm 0.1$~mag derived in the light curve analysis (Sect.~\ref{sec:analysis_phot}). The best match and two other examples with different applied host galaxy extinction values are shown in Fig.~\ref{fig:comp_spect} illustrating the extinction effects and the similarity between SNe~2005at and 2007gr. We also carried out a comparison between the spectrum of SN~2005at and the SN~2007gr spectrum $\sim$11 days after the peak, however, the wavelength range of the comparison is more limited and it also yielded a poorer fit. It is also possible that our estimate for the date of the \textit{R}-band peak of SN~2005at is slightly later than the actual \textit{R}-band maximum.

\begin{figure*}
\includegraphics[width=\linewidth]{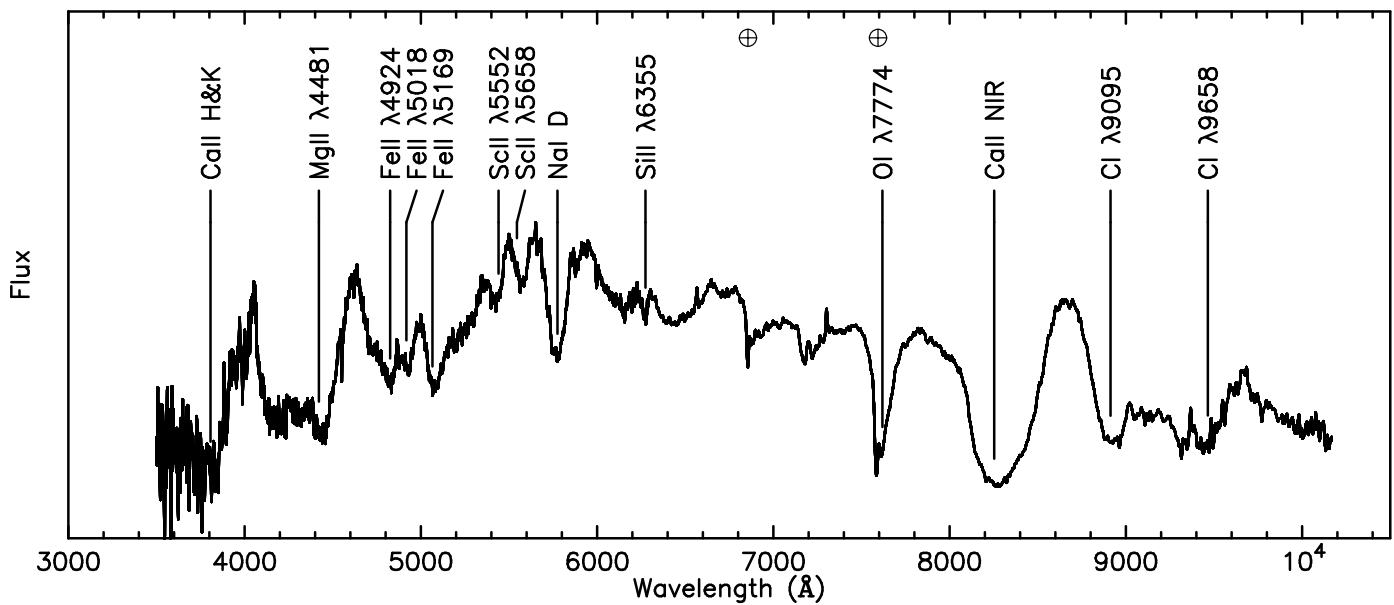}
\caption{Most prominent lines in the spectrum of SN~2005at identified. The wavelengths of the most prominent telluric features are indicated with a $\oplus$ symbol.}
\label{fig:spect_lines}
\end{figure*}

The striking similarity between the spectra of SNe~2005at and 2007gr allowed for the adoption of the line identifications from the comprehensive studies of SN~2007gr. In Fig.~\ref{fig:spect_lines} the main features in the spectrum of SN~2005at are labelled, guided by the line identifications of \citet{valenti08} and \citet{hunter09}. \citet{valenti08} noted particularly strong carbon features in the spectra of SN~2007gr compared to other type Ic SNe. The spectrum of SN~2005at does not show obvious \CII\ features at 6580 and 7235 \AA, though these lines disappeared in SN~2007gr close to maximum light. However, overlay of the spectra of SNe~2005at and 2007gr shown in Fig.~\ref{fig:comp_spect} allows detailed comparison of the \CI\ features at 9095, 9406 and 9658~\AA. One of the main differences between the two very similar spectra are the slightly weaker \CI\ features, though the \CI\ line at 9658~\AA\ is close to identical. However, we note that telluric absorption features can also have an effect on the line profiles in this region of the spectrum. 

Comparison of other lines shows that the two SNe have very similar \CaII\ NIR triplet and \OI\ $\lambda$7774 features, however the latter line is likely affected by telluric absorption. \citet{crockett08} noted the presence of narrow P Cygni lines as a distinctive characteristic of SN~2007gr, absent in many other type Ib/c SNe, however, the spectrum of SN~2005at also shows very similar lines of e.g. \FeII\ $\lambda\lambda$4924, 5018, 5169 and \ScII $\lambda\lambda$5552, 5658. Also the \SiII\ $\lambda$6355 appears to be still present in the spectrum. We calculated the velocities of the spectral features from the blueshifted positions of the absorption minima. Similar to SN~2007gr roughly two weeks after the optical maximum, the majority of the P Cygni features suggest a velocity of $\sim$6000~km~s$^{-1}$, with the exception of the \CaII\ lines arising from the higher velocity outer ejecta at $\sim$11000~km~s$^{-1}$, and \SiII\ $\lambda$6355 and \MgII\ $\lambda$4481 features tracing the photospheric expansion velocity of $\sim$4000~km~s$^{-1}$.

\subsection{Radio data}

\begin{figure}
\includegraphics[width=\linewidth]{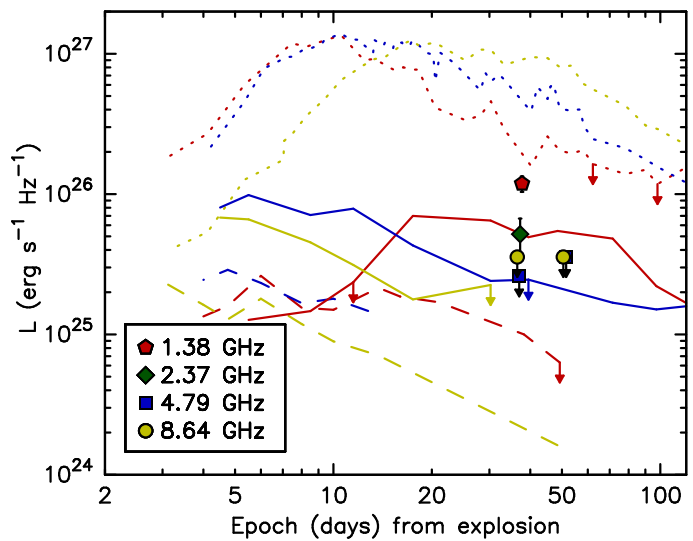}
\caption{Radio observations of SN~2005at compared to the radio evolution of type Ic SNe~1994I \citep[dotted lines;][]{weiler11}, 2002ap \citep[dashed lines;][]{berger02} and 2007gr \citep[solid lines;][]{soderberg10} at 1.4 (red), 4.9 (blue) and 8.5~GHz (yellow). Upper limits are indicated with arrows. The epoch is set from the estimated explosion date.}
\label{fig:radio}
\end{figure}

SN~2005at appears to be less luminous in the radio than other type Ib/c SNe with well-sampled radio data \citep[see][and their Fig. 2]{salas13}, with the exceptions of SNe~2007gr and 2002ap. In Fig.~\ref{fig:radio} radio data of SN~2005at are compared to the evolution of a few type Ic SNe. SN~2007gr, which shares most similar optical evolution to SN~2005at, peaked in the radio very early ($t<10$~d) at 10$^{26}$~erg~s$^{-1}$~Hz$^{-1}$ at $\sim$5~GHz, followed by a rapid decline \citep{soderberg10}. The observations of SN~2005at are consistent with a similar fast radio evolution. The SN was likely already past the radio peak and in its optically thin phase at 2.37, 4.79, and 8.64~GHz during the first epoch of observations (JD~$=2453460$), $\sim$1.2 months from the explosion. At 1.38~GHz the SN was likely close to the peak, perhaps already in its optically thin phase. The two-point spectral index $\alpha $ between 1.38 and 2.37~GHz is steep ($\sim-1.5^{+0.7}_{-0.9}$, assuming $S_{\nu} \sim \nu^{\alpha}$, where $S$ is the flux density and $\nu$ is the frequency) but this might not be representative of the optically thin part, as the 1.38~GHz emission could still be affected by either synchrotron self absorption or free-free absorption. Additionally, we note that a proper estimate of the spectral index requires the use of flux densities from maps with matched resolutions, and owing to the use of only four antennas in the first radio epoch, we have a very limited uv-dataset that would allow such an exercise. Consistent with the decline in flux at higher frequencies, the follow-up yielded only upper limits at 4.79 and 8.64~GHz $\sim$1.7 months after the explosion. 

\section{Late-time observations}
\label{sect:late-time}

\subsection{Mid-infrared images}
\label{sect:mir}

\begin{figure}
\includegraphics[width=\linewidth]{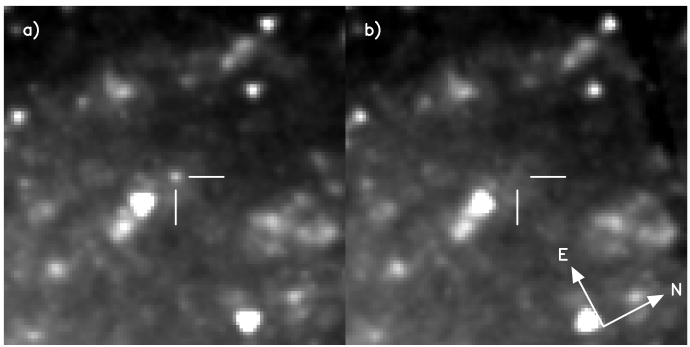}
\caption{\textit{Spitzer} 3.6~$\mu$m images of the field of SN~2005at. Observation on 2006 September 25 \textbf{(a)} shows a clear detection of a source at the location of SN~2005at, indicated with tick marks, $\sim$1.6~yr from the explosion. Non-detection was obtained on 2012 May 25.7 \textbf{(b)}, $\sim$7.3~yr from the explosion. The panels are $1 \times 1$~arcmin$^{2}$ subsections. Orientation is indicated in the \textit{second panel}.}
\label{fig:mir}
\end{figure}

The SN~2005at host galaxy NGC~6744 was observed with the \textit{Spitzer} Space Telescope using the Infrared Array Camera \citep[IRAC;][]{fazio04} on 2006 September 25.1~UT (programme 30496, PI: Fisher). Observations were carried out in all four channels (3.6, 4.5, 5.8 and 8.0~$\mu$m) and cover the explosion site of SN~2005at $\sim$1.6~yr from the explosion. NGC~6744 was re-observed by (warm) \textit{Spitzer} on 2012 May 25.7~UT using IRAC at 3.6 and 4.5~$\mu$m (programme 58871, PI: Tully). The post basic calibrated data (PBCD) image mosaics were obtained from the Spitzer Heritage Archive and the SN was identified as a clear point source present in all the 2006 images consistent with the coordinates of the SN. Panels of the 3.6~$\mu$m images are shown in Fig.~\ref{fig:mir}.

The 2012 images were aligned to match the observations from 2006 and a careful visual inspection revealed no point source apparent at the SN location. We therefore used the late-time observations at 3.6 and 4.5~$\mu$m as reference images for image subtraction to remove the strong background contamination present at the location of SN~2005at, using a slightly modified version of the {\sc isis} 2.2 package \citep{alard98, alard00}. The image subtraction package carries out a convolution of the better seeing image to match the count levels and the PSF to that of the worse seeing image. Aperture photometry with the Starlink package \textit{Gaia}\footnote{\urlwofont{http://star-www.dur.ac.uk/~pdraper/gaia/gaia.html}} was then performed in the subtracted images. In the 5.8 and 8.0~$\mu$m images for which no reference images were available we used the SNOOPY\footnote{SNOOPY, originally presented in \citet{patat96}, has been implemented in IRAF by E. Cappellaro. The package is based on daophot, but optimised for SN magnitude measurements.} PSF fitting package to measure the SN flux in the unsubtracted images. For this the SN position was fixed to the coordinates determined from the subtracted images. Aperture corrections were applied to the measured fluxes following the IRAC Instrument Handbook (version 2.0.3). At 3.6 and 4.5~$\mu$m the photometric uncertainties were estimated as the standard deviation of the fluxes measured with an identical aperture as used for the SN photometry at nine positions close to the SN position in the subtracted images. At 5.8 and 8.0~$\mu$m the photometric uncertainties were estimated by simulating and PSF-fitting nine point sources in the images close to the SN position. The mid-infrared (mid-IR) SN magnitudes are reported in Table~\ref{table:mid-ir}. 

\begin{table*}
\caption{\textit{Spitzer}/IRAC mid-IR photometry for SN~2005at.}
\centering
\begin{tabular}{ccccccccc}
\hline
\hline
JD & \multicolumn{2}{c}{3.6~$\mu$m} & \multicolumn{2}{c}{4.5~$\mu$m} & \multicolumn{2}{c}{5.8~$\mu$m} & \multicolumn{2}{c}{8.0~$\mu$m} \\
(2400000+) & (mJy) & (mag) & (mJy) & (mag) & (mJy) & (mag) & (mJy) & (mag) \\ 
\hline
54003.57  & 0.049(0.005) & 16.88(0.11) & 0.064(0.009) & 16.12(0.15) & 0.13(0.02) & 14.86(0.20) & 0.35(0.09) & 13.13(0.27) \\
\hline
\end{tabular}
\tablefoot{The errors are given in brackets.}
\label{table:mid-ir}
\end{table*}

Whereas the mid-IR evolution of several nearby type IIP SNe \citep[e.g.][]{kotak09, meikle11} has been obtained with \textit{Spitzer}, similar observations of normal type Ib/c SNe have not been extensively carried out. We note that \citet{kochanek11} reported the early mid-IR photometry of SN~2007gr, obtained a couple of weeks after the optical maximum, and found the observations to be consistent with 5000~K black body emission. However, no late-time mid-IR observations of SN~2007gr are available for comparison with SN~2005at. To our knowledge the \textit{Spitzer} observations of SN~2005at reported here, are the only late-time ($> 1$~yr) mid-IR data of a type Ib/c SN published to date. 

\begin{figure}
\includegraphics[width=\linewidth]{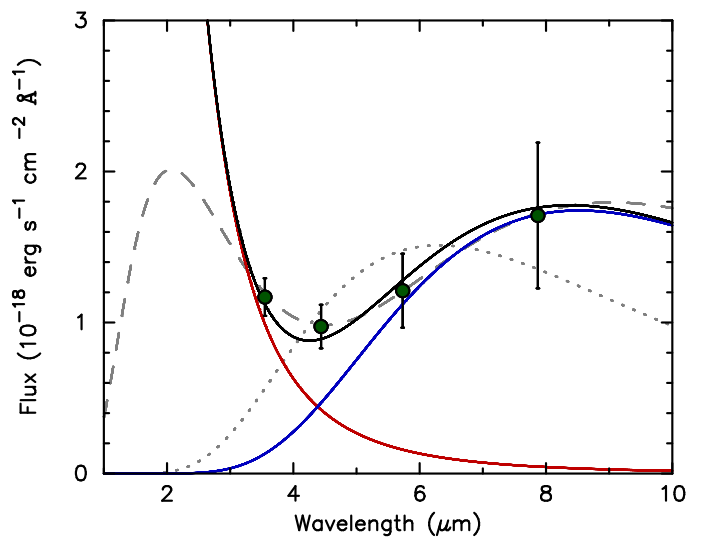}
\caption{Two-component black-body fit (black line) to the mid-IR \textit{Spitzer} observations (circles) at $\sim$1.6~yr with a 10000~K (red line) hot light echo and 340~K (blue line) cool dust components. The dashed grey line shows the fit without fixing the temperature of the hot component. The dotted grey line shows a single component fit.}
\label{fig:mid-ir}
\end{figure}

The observed mid-IR spectral energy distribution (SED) of SN~2005at cannot be reasonably fitted with a single black body (see Fig.~\ref{fig:mid-ir}). The high 3.6~$\mu$m flux cannot be explained by molecular emission. For example CO fundamental band emission, prominent in some CCSNe, can be identified as an excess at 4.5~$\mu$m \citep[see][]{kotak06}, however, no such feature is present in SN~2005at. Therefore, we explore a two-component model which, with four free parameters, yields for the hot component $T_{\mathrm{hot}}=1400$~K and $R_{\mathrm{hot}}=5.0\times10^{14}$~cm which is not physically reasonable and likely reflects overfitting of the data to the Rayleigh–Jeans tail of the black body. However, the SED fit suggests the possibility that the mid-IR emission is arising from both a hot interstellar light echo component and a cool dust associated component. To avoid overfitting the data, we adopt a canonical CCSN peak luminosity colour temperature of $T_{\mathrm{hot}}=10000$~K for the presumed hot light echo component. The effect of this on the temperature of the cool component is $\sim$10 per cent compared to the fit with four free parameters and yields $T_{\mathrm{cool}}=340$~K and $R_{\mathrm{cool}}=1.6\times10^{16}$~cm. The cool 340~K component is consistent with dust formation in the ejecta expanding at $\sim$3000~km~s$^{-1}$. However, we cannot fully rule out contamination by an unrelated background/foreground source to SN 2005at fluxes in the 5.8$-$8~$\mu$m range since we do not have reference images at these wavelengths \citep[see e.g. \textit{Spitzer} observations of SN 2002hh;][]{meikle06}.

Mid-IR observations of CCSNe have previously revealed thermal IR echoes, from e.g. SNe~2003gd \citep{meikle07}, 2004et \citep{kotak09}, and 2004dj \citep{meikle11}. However, this might be the first case of a non-thermal interstellar light echo observed at mid-IR. Furthermore, unlike with nearby type II SNe, thermal IR echoes from pre-existing dust or dust formation in the ejecta, are not commonly observed features in normal type Ib/c SNe, with some rare exceptions such as SNe~1982E and 1982R \citep{graham86} and 1990I \citep{elmhamdi04}. In the case of the peculiar Ibn SN~2006jc the near- and mid-IR properties were explained as a combination of newly-formed and pre-existing dust in the circumstellar medium \citep[e.g.][]{mattila08}. However, the rarity of nearby type Ib/c SNe and the lack of suitable observations might also explain in part why light echo and dust formation observations in type Ib/c SNe are not commonly reported. In the follow-up of SN~2007gr, \citet{hunter09} found no evidence of dust formation, based on the near-IR light curve and colour evolution and optical spectra, out to $\sim$1.1~yr. However, their extensive follow-up did not cover a comparable epoch to that probed by the \textit{Spitzer} data of SN~2005at.

\subsection{HST post-explosion images}
\label{sect:hst}

The HST archive contains multiple epochs of observations of NGC~6744, however, none of the pre-explosion images obtained with multiple different instruments onboard the HST cover the explosion site of SN~2005at, as already noted by \citet{smartt09b}. However, the explosion site of SN~2005at was observed with the HST + Wide-Field and Planetary Camera 2 (WFPC2) on 2007 April 13 (snapshot programme 10877, PI: Li), $\sim$2.1 years after discovery. $2 \times 230$~s exposures were taken in the \textit{F555W} filter, and $2 \times 350$~s in \textit{F814W}. We aligned the mosaiced WFPC2 \textit{F555W} image (file: hst\_10877\_04\_wfpc2\_f555w\_wf\_drz.fits, downloaded from the Hubble Legacy Archive) to the DFOSC \textit{V}-band image of SN~2005at taken on 2005 April 19. The alignment was carried out using the WFPC2 mosaic, instead of just the PC chip image on which SN~2005at lies, as the latter has too small a field of view to be matched against the DFOSC image. 25 sources common to both frames were identified, their pixel coordinates were measured in both frames, and a geometric transformation between the two was derived using IRAF {\sc geomap}, with an rms error in the fit of 50~mas. We then aligned the WFPC2 mosaic (which has been resampled to a uniform pixel scale of 0.1\arcsec/pix) to the PC chip image on which we performed photometry. The same technique as before was used, with an rms error in the alignment of 15~mas. We can hence locate the position of SN~2005at on the PC chip to 52~mas, or $\sim$1 PC chip pixel.

\begin{figure*}
\centering
  \includegraphics[width=0.32\linewidth]{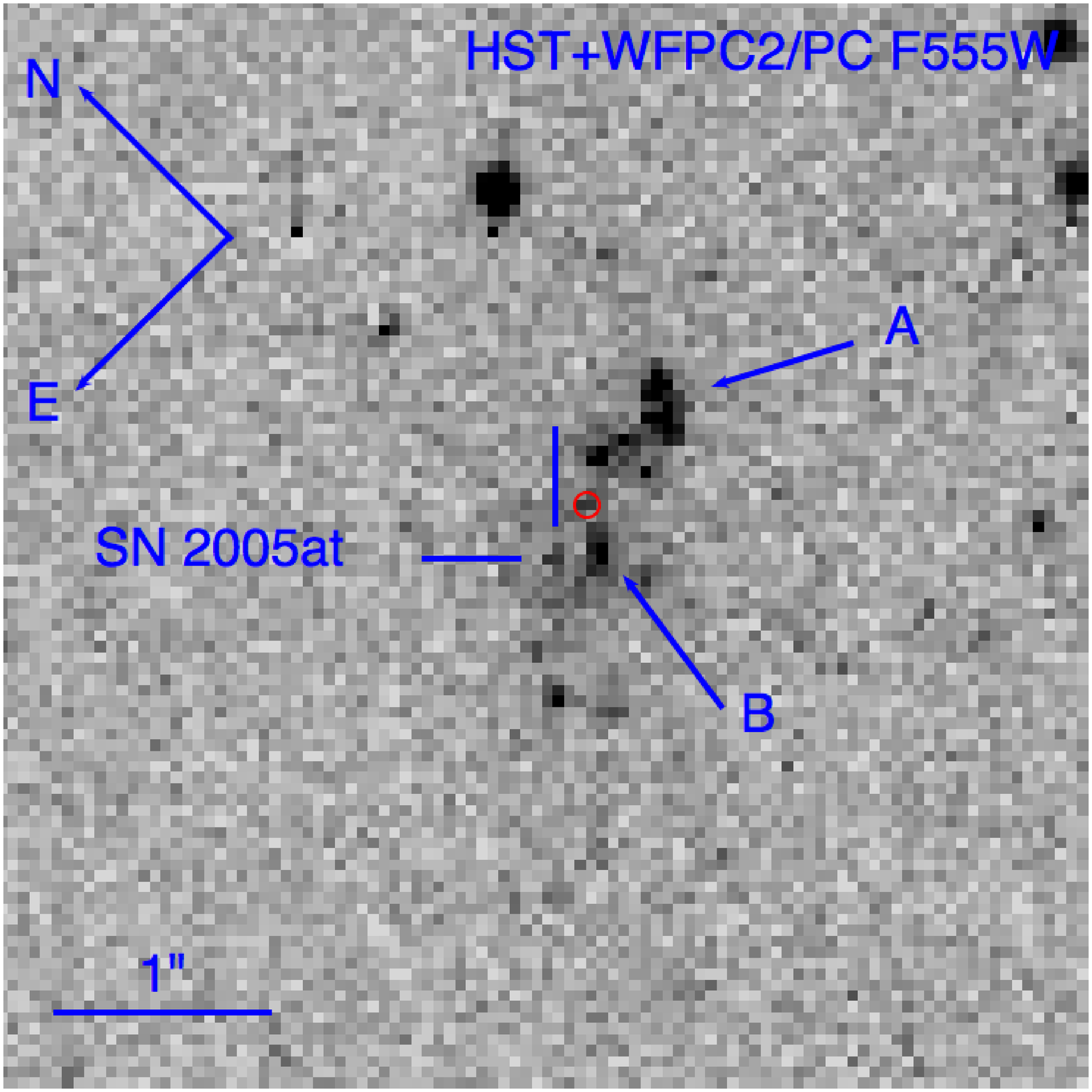}
     \includegraphics[width=0.32\linewidth]{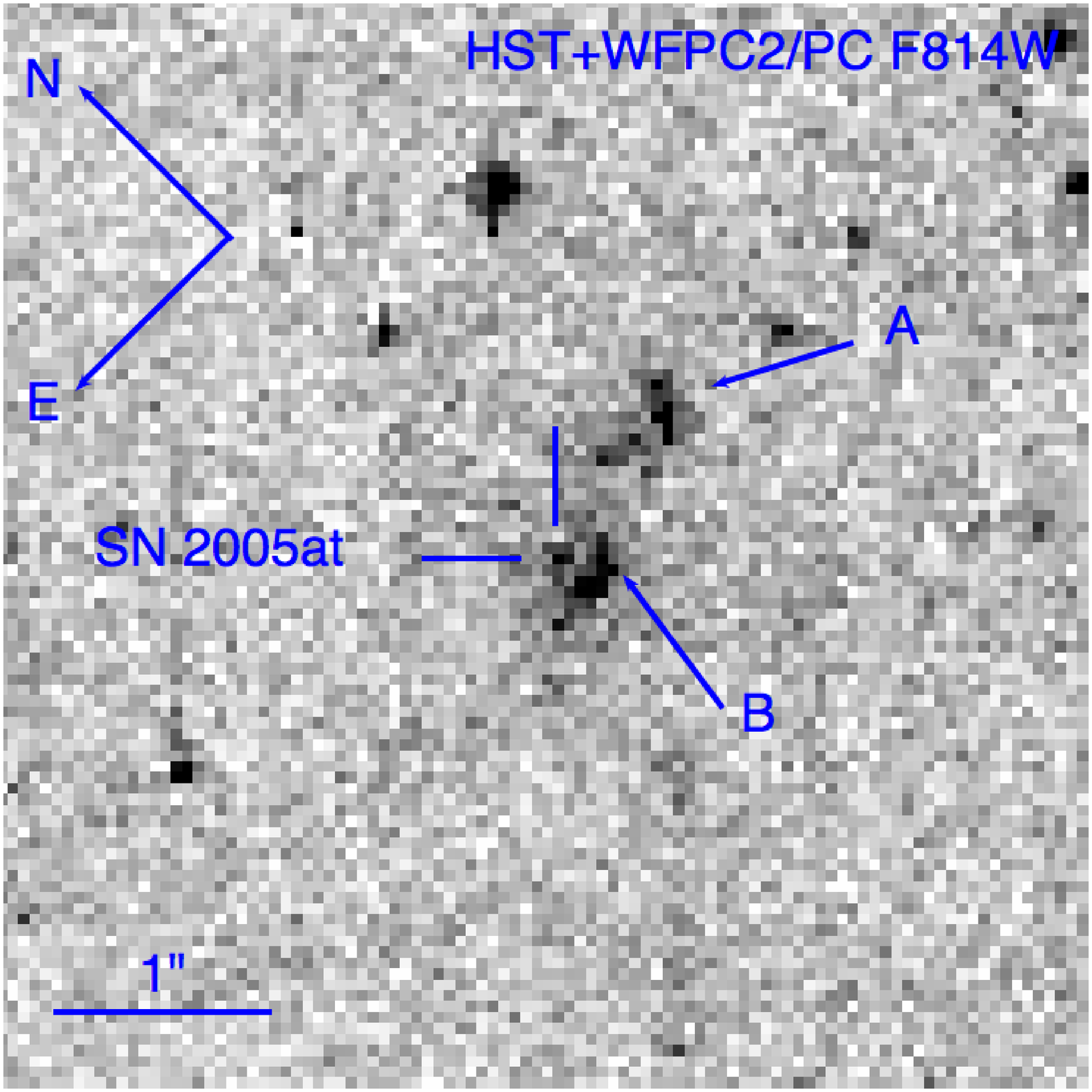}
     \includegraphics[width=0.32\linewidth]{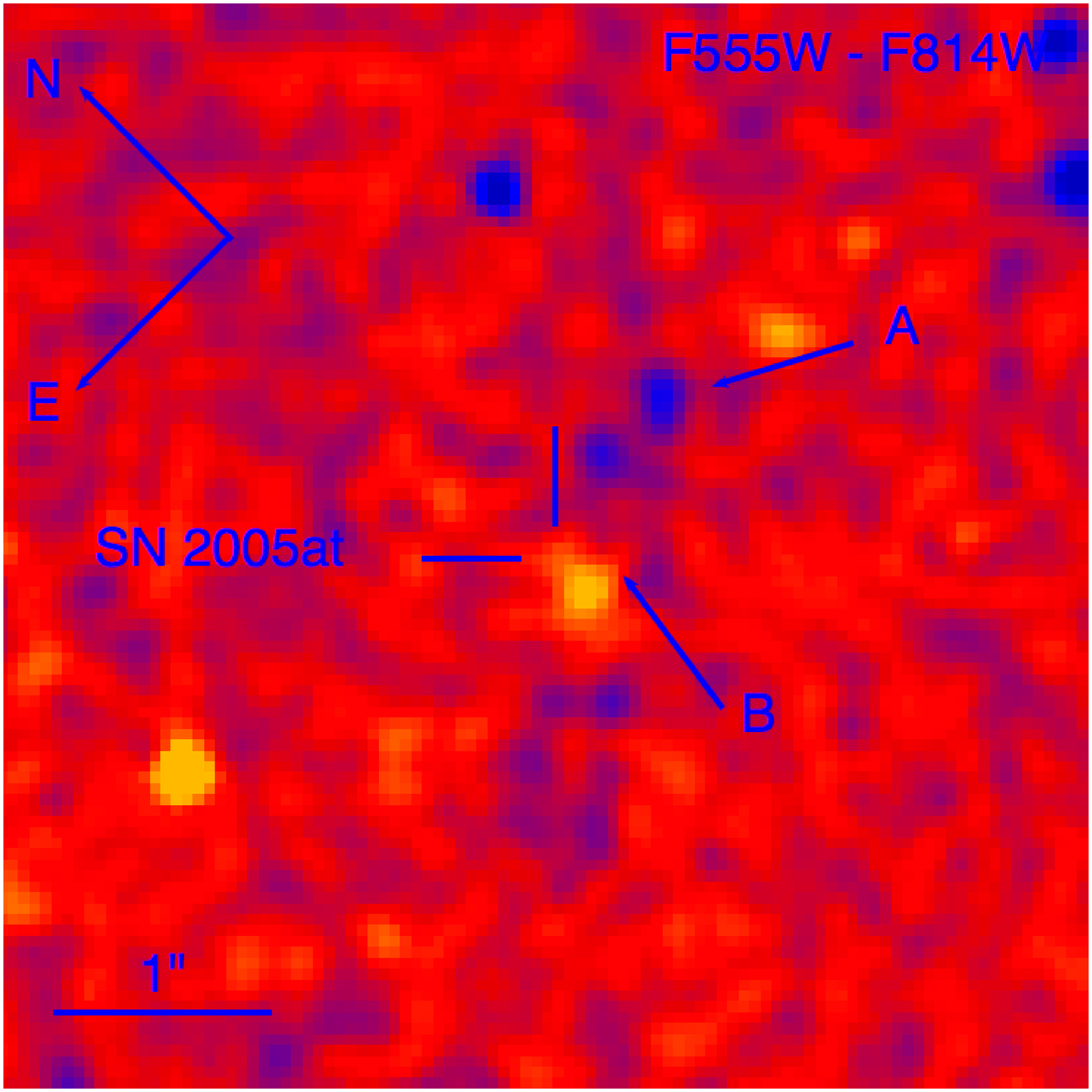}
\caption[]{HST WFPC2/PC \textit{F555W} \textbf{a)} and \textit{F814W} \textbf{b)} images of the site of SN~2005at. The source which we associate with the SN is indicated by the tick marks, the extended sources A and B are also shown. The red circle shows the transformed position of the \HeII\ source (Sect.~\ref{sect:heii}). The \textit{F555W}$-$\textit{F814W} \textbf{c)} image has been Gaussian smoothed for illustration purposes. The different colours of sources A and B are evident. The scale and orientation of all three panels are identical and indicated in the panels on the right.}
\label{fig:wfpc2}
\end{figure*}

We find a point-like source coincident with the transformed position of SN~2005at on the WFPC2 \textit{F555W} image, see Fig.~\ref{fig:wfpc2}. We used the {\sc HSTphot} package \citep{dolphin00} to carry out PSF fitting photometry on this source. The source is detected in both \textit{F555W} and \textit{F814W} at a combined significance of 10$\sigma$, and has $\chi = 1.31$, a sharpness of $-0.29$ and an object type classification from {\sc HSTphot} which are consistent with a point source, which should have $\chi < 1.5$ and a sharpness parameter between $-0.3$ and $+0.3$. We measure magnitudes of $m_{F555W} = 24.62 \pm 0.16$~mag and $m_{F814W} = 23.67 \pm 0.14$~mag in the HST Vegamag system, which correspond to Johnson-Cousins magnitudes of $m_{V} = 24.59 \pm 0.16$~mag and $m_{I} = 23.64 \pm 0.14$~mag. The identity of the source is ambiguous, with several plausible possibilities:

\begin{itemize}

\item {\it The source is SN~2005at.} The \textit{V}-band magnitude, $m_{V}=24.62$~mag, is broadly consistent with the decline rate of 0.0098~mag~d$^{-1}$ expected from the radioactive decay of $^{56}$Co, and starting close in time to the SALT observations (171 days from the explosion). However, this would imply that the ejecta is still opaque to $\gamma$-rays after nearly 2 years. Only 1-2 M$_ {\sun}$ of ejecta was estimated for SN~2007gr \citep{hunter09,mazzali10}, and assuming the same ejecta mass it is implausible that SN~2005at would be fully trapped this long. In fact the observed \textit{V}-band decline rate of SN~2005at in the early tail phase is much steeper and extrapolation of the light curve would suggest SN~2005at to be several magnitudes fainter than the observed source. Therefore, we find it very unlikely that the source is SN~2005at.

\item {\it The source is a binary companion to SN~2005at.} If the source is a binary companion to SN~2005at, then it should suffer from the same reddening as the SN. The source would then be intrinsically extremely luminous and consistent for example only with the most luminous Wolf-Rayet stars \citep{massey02, eldridge13}.

\item {\it The source is a light echo of SN~2005at.} Given that SN~2005at suffers from a relatively high level of extinction, it is plausible that this dust could give rise to an unresolved light echo at $t\sim2$ years, as seen e.g. in the case of SN~2007od and SN~2007it \citep{andrews10, andrews11}. In this scenario the red colour of the source is due to the colour of SN~2005at at peak, which will contribute most of the flux to the light echo. Assuming a FWHM of a point source is $\sim$1.5 PC chip pixels, we only resolve sources larger than $b \sim 3.5$~pc at the distance of SN~2005at. We can estimate the maximum distance $l$ of the dust cloud from the SN giving rise to the possible light echo \citep[see e.g.][]{liu03} to be

\begin{equation} \label{eq:cloud}
      l \lesssim \frac{b^{2} - (ct)^{2}}{2 ct} = 10 \mathrm{pc} \,.
\end{equation} 

Future high-resolution observations would assist substantiating the light echo scenario since we would expect a light echo to have changed in luminosity by now. \textit{Spitzer} observations of SN~2005at at $\sim$1.6~yr (Sect.~\ref{sect:mir}) are also consistent with a light echo.

\item {\it The source is an unresolved cluster.} We cannot exclude the source from being a compact cluster on the basis of the source being point-like. For example a cluster like G293 \citep{williams01} in M31 with $V \sim -7$~mag and $r \sim 3$~pc would be consistent with our observations and has an age of $\sim$60~Myr. Unfortunately the $V-I$ colour is a poor diagnostic of cluster age \citep[see][for an example]{valenti11}, however, with future observations a clear test between these two most likely scenarios could be carried out. Also, with possible HST \textit{U} and \textit{B}-band photometry, the age of the cluster could be accurately constrained (as these filters probe the young and hot stellar population).

\item {\it An unrelated source.} We note that the observations do not allow us to completely exclude an unrelated source located on the line-of-sight by chance.

\end{itemize}

Within 1\arcsec\ of the position of SN~2005at are two extended sources, shown in Fig.~\ref{fig:wfpc2}. To the west of SN~2005at lies the more distant source A, which appears relatively blue. Source B is closer ($\lesssim$~0.2\arcsec, $\sim$10~pc) to the position of SN~2005at, but appears to have a somewhat different morphology in the \textit{F555W} and \textit{F814W} filters. The \textit{F555W} flux is concentrated to the northwest of the source, while the \textit{F814W} flux lies more to the southeast. The colour difference across source B (see \textit{F555W}$-$\textit{F814W} colour map in Fig.~\ref{fig:wfpc2}) makes us wary of attempting to fit a model stellar population to the photometry, as it is most likely not a single, coeval cluster.

We also searched all publicly accessible data archives of ground based telescopes for further imaging which may be of use in constraining the progenitor of SN~2005at. Unfortunately, as the images available are of much poorer resolution than the HST data (sources A and B were blended), we were unable to use them to further constrain the progenitor environment. 

\section{Explosion site of SN~2005at}
\label{sect:site}

\subsection{He II post-explosion images}
\label{sect:heii}

\citet{bibby13} reported their initial results from a study finding a majority of the Wolf-Rayet star candidates in the host galaxy of SN~2005at, NGC~6744, to be associated with \HII\ regions. Their study was based on the narrowband \HeII\ $\lambda$4684 on-filter and $\lambda$4781 off-filter observations of NGC~6744 obtained on 2008 July 4 with the Very Large Telescope (VLT) using the FOcal Reducer and low dispersion Spectrograph 1 \citep[FORS1;][]{appenzeller98}. The \HeII\ $\lambda$4684 filter covers the strong emission lines of \HeII\ $\lambda$4686 and \CIII\ $\lambda$4650 arising from nitrogen (WN) and carbon (WC) sequence Wolf-Rayet stars, respectively \citep{crowther07}. 

\begin{figure}
\includegraphics[width=\linewidth]{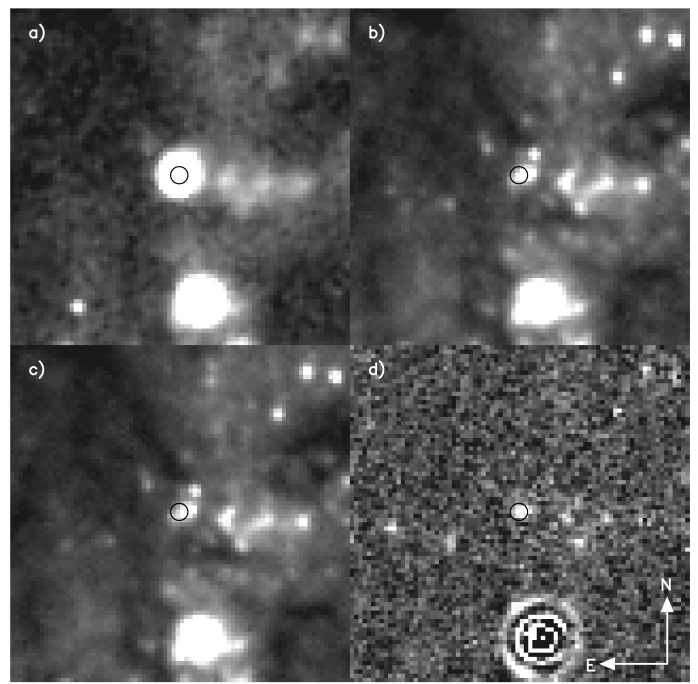}
\caption{\textbf{a)} DFOSC \textit{B}-band image of SN~2005at obtained on 2005 April 19. \textbf{b)} FORS1 narrowband on-filter and \textbf{c)} off-filter images, centered at 4684~\AA\ and 4781~\AA\ respectively, of the explosion site of SN~2005at obtained on 2008 July 4. \textbf{d)} Subtraction between the on and off-filters revealing a source of \HeII\ $\lambda$4686 and/or \CIII\ $\lambda$4650 line emission coincident with the explosion site of SN~2005at. All panels are $20 \times 20$~arcsec$^{2}$ subsections of the images. The location of the SN is shown with a circle with a diameter of 1~arcsec. North is up and east is to the left in all the panels.}
\label{fig:heii}
\end{figure}

We downloaded these raw narrowband images of NGC~6744 from the European Southern Observatory (ESO) Archive, carried out standard bias and flat-field corrections, and aligned individual frames. The final combined on and off-filter images were subtracted from each other using the {\sc isis} 2.2 package to remove the underlying continuum level from the total emission observed with the on-filter. To spatially compare the subtracted image with the position of SN~2005at, the DFOSC \textit{B}-band image (pixel scale 0.39\arcsec/pix) of SN~2005at obtained on 2005 April 19 was aligned to match the \HeII\ $\lambda$4684 image (pixel scale 0.25\arcsec/pix). 20 stars common in both images were used to derive a general geometric transformation using the {\sc geomap} and {\sc geotran} tasks in IRAF, with a rms error of 30~mas. The result is shown in Fig.~\ref{fig:heii}. SN~2005at is found to be close to a source appearing slightly extended in the continuum subtracted \HeII\ image. The offset between SN~2005at and the centred position of the line emission source is measured to be roughly 0.22\arcsec, i.e. less than a pixel. The observed line emission is likely arising from a cluster or clusters which host Wolf-Rayet stars with prominent helium and carbon lines. Such stars are expected to be progenitors of at least some stripped-envelope SNe. We cannot exclude SN~2005at from being associated with this region. We note that the narrowband images have been obtained in such a late stage that any contribution from line emission arising from the SN would be negligible.

We also aligned the FORS1 \HeII\ $\lambda$4684 on-filter image to the WFPC2/PC \textit{F555W} image (Sect.~\ref{sect:hst}). 12 sources were identified in common to both frames, and an rscale transformation was derived between the two frames with an rms error of 60~mas (1.2 PC chip pixels). We find the emission as measured in the continuum subtracted \HeII\ image to be centred on a position between source A and B, approximately 0.25\arcsec\ from SN~2005at. The \HeII\ flux is coincident with a source detected by {\sc HSTphot}, and classified as extended with a magnitude and colour of $m_{V} = 24.21 \pm 0.01$, $V-I = 0.21 \pm 0.01$.

\subsection{Ground-based pre-explosion images}

For completeness, we also analysed available deep pre-explosion ground-based images of NGC~6744 covering the explosion site of SN~2005at, where there were corresponding deep late-time archive images in similar filters. In the optical these include ESO 2.2-m telescope +  Wide Field Imager (WFI) 3~$\times$~900~s \textit{B} and 3~$\times$~600~s \textit{V}-band images obtained on 1999 August 28 (pixel scale 0.238\arcsec/pix, $FWHM \sim 1.3$\arcsec). In the near-IR we made use of the VLT Infrared Spectrometer And Array Camera (ISAAC) 6~$\times$~8~$\times$~15~s exposures obtained in \textit{Ks}-band on 2003 June 24 (pixel scale 0.1484\arcsec/pix, $FWHM \sim 0.9$\arcsec). 

The late-time optical reference images were obtained with the VLT + FORS1 on 2008 June 30 with 2~$\times$~250~s and 1~$\times$~250~s exposures in \textit{B} and in \textit{V}, respectively. The \textit{Ks}-band reference image of NGC~6744 was obtained with the ESO 3.58-m New Technology Telescope (NTT) + Son of ISAAC (SofI) on 2012 March 14 and was combined with 32~$\times$~12~$\times$~7~s exposures. 

All the images were reduced using standard IRAF based techniques. In all the cases the reference images had superior seeing compared to the pre-explosion images. Aligned reference images were subtracted from the precursor images using the {\sc isis} 2.2 package. This resulted in smooth subtractions at the explosion site of SN~2005at in all the studied wavebands, and showed no signs of a disappearing source. We derived 5$\sigma$ detection upper limits for each subtracted image pair yielding $m_{B} > 23.3$~mag, $m_{V} > 22.8$~mag, and $m_{K} > 20.1$~mag. Taking into account the distance modulus and the derived total line-of-sight extinction of $A_{V} = 2.0$~mag this converts into absolute magnitude limits of $M_{B} > -9.3$~mag, $M_{V} > -9.1$~mag, and $M_{K} > -10.0$~mag. Unfortunately these limits are too shallow to provide any meaningful constraints on the progenitor of SN~2005at. In good agreement with the initial estimates of \citet{eldridge13}, these upper limits of SN~2005at are at least $\sim$1.5~mag brighter than the most luminous Wolf-Rayet stars in the Large Magellanic Cloud \citep[see their Fig.~8 and also][]{massey02}. Compared to other events presented in the literature, more constraining upper limits have been derived for a selection of type Ib/c SN progenitors, including SN~2007gr with $M_{B} > -6.7$~mag and $M_{I} > -8.6$~mag \citep[see][]{crockett08, eldridge13}.

\section{Conclusions}
\label{sect:conclusions}

With the multi-band light-curve comparison we derive a host galaxy line-of-sight extinction of $A_{V} = 1.9 \pm 0.1$~mag for SN~2005at at 9.5~Mpc. Based on the available photometric data and assuming the above mentioned extinction, we conclude that SN~2005at is very similar to SN~2007gr. This is evident by the excellent overall match between the optical light curves and spectroscopy of SNe~2005at and 2007gr. The very small (0.1~mag) inferred difference between the absolute optical magnitudes of the two SNe is well within the errors of host galaxy distance estimates. The optical spectroscopy of both SNe~2005at and 2007gr shows similar and unusually narrow spectral line features. SN~2005at appears to be less luminous at radio wavelengths than the majority of other type Ib/c SNe, which was also the case with SN~2007gr. SN~2005at likely had very similar explosion properties to those of SN~2007gr and the occurrence of another nearby event which shares such strong similarity to the well-monitored SN~2007gr also underlines the normal type Ic template nature of SN~2007gr. 

We recover SN~2005at in four mid-IR bands of \textit{Spitzer} archive images $\sim$1.6~yr post-explosion and find the emission consistent with a combination of a light echo and newly-formed dust in the ejecta. The study of the late-time high-resolution HST images revealed a faint point source coincident with the explosion site of SN~2005at $\sim$2.1~yr after the SN explosion. We find the unresolved source to be most likely either a declining light echo of the SN or a compact cluster. However, we cannot fully exclude an extremely luminous companion star of the SN~2005at progenitor, or an unrelated object as being the source of the late-time emission. Based on ground-based narrow-band imaging, we find the SN~2005at explosion site to overlap with an extended region of observed line emission at 4684~\AA, consistent with helium and/or carbon emission of Wolf-Rayet stars. Analysis of the ground-based pre-explosion images yielded only shallow upper limits for the SN~2005at progenitor star and could not be used to constrain the nature of the progenitor. 

Detailed studies of nearby and reddened SNe -- both categories into which SN~2005at falls -- are encouraged due to the importance of these SNe for estimating extinction and missing-fraction corrections for studies of SN rates. Furthermore, various statistical studies of SNe aim to use local volume limited SN samples which are as complete as possible.

\begin{acknowledgements}

We thank the referee Stacey Habergham for very useful comments. We thank Brian Schmidt for providing the spectrum of SN~2005at. We thank Harry Lehto, Zsolt Paragi, Laura Portinari and Rami Rekola for very helpful discussions and suggestions. E.K. acknowledges financial support from the Academy of Finland (project: 263854) and the Jenny and Antti Wihuri Foundation. This work was partly supported by the European Union FP7 programme through ERC grant number 320360. C.R.C. acknowledges financial support from the ALMA-CONICYT FUND Project 31100004. P.L. acknowledges support from the ERC-StG grant EGGS-278202. The Dark Cosmology Centre is funded by DNRF. This research has made use of the NASA/IPAC Extragalactic Database (NED) which is operated by the Jet Propulsion Laboratory, California Institute of Technology, under contract with the National Aeronautics and Space Administration. This work is based in part on observations made with the \textit{Spitzer} Space Telescope, which is operated by the Jet Propulsion Laboratory, California Institute of Technology under a contract with NASA. This research has made use of the NASA/IPAC Infrared Science Archive, which is operated by the Jet Propulsion Laboratory, California Institute of Technology, under contract with the National Aeronautics and Space Administration. Some of the data presented in this paper were obtained from the Mikulski Archive for Space Telescopes (MAST). STScI is operated by the Association of Universities for Research in Astronomy, Inc., under NASA contract NAS5-26555. Support for MAST for non-HST data is provided by the NASA Office of Space Science via grant NNX13AC07G and by other grants and contracts. Some of the observations reported in this paper were obtained with the Southern African Large Telescope (SALT). The Australia Telescope Compact Array is part of the Australia Telescope National Facility which is funded by the Commonwealth of Australia for operation as a National Facility managed by CSIRO. Based on observations made with ESO Telescopes at the La Silla Paranal Observatory under programme IDs 081.B-0289, 71.D-0235 and 184.D-1140 (within the European supernova collaboration led by Stefano Benetti). We have made use of the Weizmann interactive supernova data repository (\urlwofont{www.weizmann.ac.il/astrophysics/wiserep}).

\end{acknowledgements}


\end{document}